# Electronic band structure of $Ti_2O_3$ thin films studied by angle-resolved photoemission spectroscopy

Naoto Hasegawa[1], Kohei Yoshimatsu[1,2,*], Daisuke Shiga[1], Tatsuhiko Kanda[1], Satoru Miyazaki[1], Miho Kitamura[3], Koji Horiba[3], and Hiroshi Kumigashira[1,2,3]

[1] Institute of Multidisciplinary Research for Advanced Materials (IMRAM), Tohoku University, Sendai, 980–8577, Japan

[2] Materials Research Center for Element Strategy (MCES), Tokyo Institute of Technology, Yokohama 226–8503, Japan

[3] Photon Factory, Institute of Materials Structure Science, High Energy Accelerator Research Organization (KEK), Tsukuba, 305–0801, Japan

## Abstract

$Ti_2O_3$ exhibits a unique metal–insulator transition (MIT) at approximately 450 K over a wide temperature range of $\sim 150$ K.    This broad MIT accompanied by lattice deformation differs from the sharp MITs observed in most other transition-metal oxides.    A long-standing issue is determining the role of electron–electron correlation in the electronic structure and MIT of $Ti_2O_3$.    However, the lack of information about the band structure of $Ti_2O_3$ has hindered investigating the origin of its unusual physical properties.    Here, we report the electronic band structure of "insulating" $Ti_2O_3$ films with slight hole doping by angle-resolved photoemission spectroscopy (ARPES).    ARPES showed clear band dispersion on the surface of single-crystalline epitaxial films.    The experimentally obtained band structures were compared with band-structure calculation results based on density functional theory (DFT) with generalized gradient approximation $+ U$ correction.    The obtained band structures are in good agreement with the DFT calculations at $U = 2.2$ eV, suggesting that electron–electron correlation plays an important role in the electronic structure of $Ti_2O_3$.    Furthermore, the detailed analyses with varying $U$ suggest that the origin of the characteristic MIT in $Ti_2O_3$ is a semimetal–semimetal or semimetal–semiconductor transition caused by changes in the Fermi surface due to lattice deformation.

*Correspondence: kohei.yoshimatsu.c6@tohoku.ac.jp

**INTRODUCTION**

$Ti_2O_3$, which has a corundum-type crystal structure, exhibits a unique metal–insulator transition (MIT). The bulk $Ti_2O_3$ is a nonmagnetic insulator with a small bandgap energy of approximately 100 meV [1] at low temperatures, and it shows a transition to semimetallic states at temperatures of approximately 450 K [2–9], which extends over a broad temperature of ~150 K. This is a unique type of transition that is not observed in most other transition-metal oxide systems. Although the crystal symmetry remains unchanged across the MIT, the unit-cell $c/a$ ratio changes significantly, which suggests a close relationship between the MIT and lattice deformations [2–6, 9–12].

Over the past few decades, the MIT mechanism has been experimentally and theoretically investigated [1–32]. The most relevant phenomenon in this mechanism is the overlap of the $a_{1g}$ and $e_g^\pi$ bands due to the Ti–Ti distance modulations along the $c$-axis of the crystal lattice [5, 6]. Owing to the trigonal distortions, the $t_{2g}$ levels in $TiO_6$ octahedra further split into a lower non-degenerate $a_{1g}$ level and higher double-degenerate $e_g^\pi$ levels. The $a_{1g}$ orbitals between the face-shared $TiO_6$ octahedra along the $c$-axis are strongly hybridized to form $a_{1g}$ and $a_{1g}^*$ bands with $e_g^\pi$ bands between them. When the Ti–Ti bond distances along the $c$-axis are short, the energy splitting between the $a_{1g}$ and $a_{1g}^*$ bands becomes large, and consequently, the $e_g^\pi$ bands do not overlap with the $a_{1g}$ bands. Furthermore, only the $a_{1g}$ band is completely filled with Ti $3d$ electrons; therefore, $Ti_2O_3$ acts as an insulator.

The $Ti_2O_3$ energy diagram describes its electronic structures and reveals the close connection between the MIT and $c/a$ ratio. In fact, the $a$-axis and $c$-axis lattice constants of bulk $Ti_2O_3$ vary significantly with temperature, and the $c/a$ ratio increases from 2.648 to 2.701 in the range of 373 to 553 K across the MIT [3]. The modulation of Ti $3d$ electron occupations in the $a_{1g}$ orbitals associated with a change in the $c/a$ ratio was revealed from the temperature dependence of linear dichroism in Ti $2p$ X-ray absorption spectra [8, 14, 15].

However, the band-structure calculations challenge the validity of this simple phenomenological model because the $a_{1g}$ and $e_g^\pi$ bands always overlap for typical Ti–Ti distances [10]; a short Ti–Ti distance of less than 2.2 Å is required to form the aforementioned insulating band diagram, which suggests the importance of electron–electron correlations in $Ti_2O_3$. To better understand the unusual physical properties of $Ti_2O_3$, it is necessary to collect detailed information about the complicated electronic band structure of the material.

Angle-resolved photoemission spectroscopy (ARPES) is a unique and powerful experimental technique for determining the band structure of a solid and has long played a central role in studies of the electronic properties of strongly correlated electron systems [33–49]. However, there have been few ARPES studies on $Ti_2O_3$ with three-dimensional corundum-type crystal structures [19]. This is mainly due to the difficulty in obtaining single-crystal surfaces of $Ti_2O_3$ using standard surface-preparation techniques, such as cleaving or sputtering and annealing. In addition, the chemically active surface of $Ti_2O_3$ [23] restricts surface-sensitive ARPES measurements on chemically well-defined surfaces. Thus, the lack of information about the band structures near the Fermi level ($E_F$), especially the Fermi surface (FS), has limited the understanding of the physics of $Ti_2O_3$.

In this study, to elucidate the electronic structure near $E_F$ of $Ti_2O_3$, we performed soft-X-ray (SX) ARPES [42–49] on single-crystalline epitaxial $Ti_2O_3$ films grown on $\alpha$-$Al_2O_3$ substrates. We synthesized the films with slight hole doping by laser molecular beam epitaxy (MBE) to obtain an atomically flat and well-ordered surface. By using the well-defined surface of the epitaxial films, we clearly observed the electronic band structure of $Ti_2O_3$. The Ti $3d$-derived hole band forms an open triangular-pyramidal-vase-shaped FS along the Γ–A line in the hexagonal Brillouin zone (BZ). The experimental band structures were compared with the band-structure calculations based on density functional theory (DFT) with generalized gradient approximation (GGA) + $U$ correction. Detailed analysis with varying $U$ revealed that the obtained band structures are well-described by the DFT calculations at $U = 2.2$ eV, indicating that the electron–electron correlation plays an important role in the electronic structure of $Ti_2O_3$. Furthermore, the comparison between the ARPES results and DFT + $U$ calculations with varying $U$ and $c/a$ ratio suggests important implications regarding the origin of the characteristic MIT in $Ti_2O_3$; the MIT is a semimetal–semimetal or semimetal–semiconductor transition caused by changes in the FS due to lattice deformation.

**EXPERIMENTAL**

Epitaxial $Ti_2O_3$ films with a thickness of approximately 100 nm were grown on $\alpha$-$Al_2O_3$ (0001) substrates by laser MBE. A sintered $TiO_x$ pellet was used as the ablation target [50]. A Nd:$Y_3Al_5O_{12}$ (Nd:YAG) laser was used for the target ablation using its third harmonic wave ($\lambda = 355$ nm) with a fluence of approximately 1.5 J/$cm^2$ and repetition rate of 5 Hz. Prior to film growth, $\alpha$-$Al_2O_3$ substrates were annealed in air at 1100°C for 2 h to obtain an atomically flat surface with step-and-terrace structures. During the deposition, the substrate temperature was maintained at 1000°C, and the oxygen pressure was maintained at $5 \times 10^{-7}$ Torr. After deposition, the oxygen gas supply was immediately turned off, and the films were quenched to room temperature to prevent overoxidation [16, 17, 51, 52].

After growth, the films were transferred *in vacuo* to the photoemission chamber using a mobile vacuum-transfer vessel to reduce surface contamination and additional oxidation. During transportation, the samples were under an ultrahigh vacuum below $5.0 \times 10^{-10}$ Torr. Photoemission spectroscopy (PES) and ARPES were conducted at the BL-2A MUSASHI beamline of the Photon Factory, KEK. The PES spectra were recorded at 100 K using a VG-Scienta SES-2002 analyzer with total energy resolutions of 250 and 500 meV at photon energies ($hv$) of 800 and 1486 eV, respectively. In the ARPES experiments in the SX region of $hv = 300$–640 eV, the energy and angular resolutions were set to approximately 150–250 meV and 0.25°, respectively. The ARPES experiments were also conducted at 100 K in the $p$-polarization geometry [36–40]. $E_F$ was inferred from gold foil in electrical connection with the sample. The surface structure and cleanliness of the vacuum-transferred films were examined by low-energy electron diffraction (LEED) and core-level photoemission measurements immediately before the ARPES measurements. The LEED pattern of the films showed six sharp diffraction spots on a low background, as shown in Fig. 1(a), reflecting the long-range crystallinity and cleanliness of the surface. No detectable C $1s$ peak was observed in the core-level photoemission spectra (not shown). These results indicate that the $Ti_2O_3$ films have high surface crystallinity and cleanliness required for ARPES measurements.

The surface morphology of the measured $Ti_2O_3$ films was analyzed by *ex-situ* atomic force microscopy (AFM) in air, and atomically flat surfaces were observed, as shown in Fig. 1(b). The crystal structure was characterized by X-ray diffraction (XRD) (see Fig. S1 in the Supplemental Material [50]), which confirmed the epitaxial growth of single-phase $Ti_2O_3$ films on the substrates. The out-of-plane and in-plane epitaxial relationships were $Ti_2O_3$ [0001] // $\alpha$-$Al_2O_3$ [0001] and $Ti_2O_3$ [11–20] // $\alpha$-$Al_2O_3$ [11–20], respectively. The temperature dependence of the electrical resistivity was measured using the standard four-probe method. The transport properties were in good agreement with previously reported values [16] (see Fig. S2 in the Supplemental Material [50]). Detailed characterizations of the grown films are presented in the Supplemental Material [50].

DFT-based band-structure calculations were conducted using QUANTUM ESPRESSO

software [53, 54]. The Perdew–Burke–Ernzerhof generalized gradient approximation (PBE-GGA) was adopted as the exchange-correlation functional [55]. The kinetic energy (charge density) cut-off was set to 60 (600) Ry. The Ti and O atomic positions were optimized by the Monkhorst–Pack scheme using a $6 \times 6 \times 6$ *k*-point grid in a self-consistent scheme [56]. In the self-consistent calculation, the lattice parameters were fixed as the experimental values for the present film ($a = 5.102$ Å, $c = 13.80$ Å, $c/a = 2.70$), which were determined by reciprocal space mapping taken at room temperature, as shown in Fig. S1(b) in the Supplemental Material [50]. Further computing details are described in the Supplemental Material [50].

## RESULTS AND DISCUSSION

Before discussing the ARPES results, we provide evidence for the fact that the properties of $Ti_2O_3$ are retained in the surface region of the films accessible for SX-ARPES [42–49] because the surface of $Ti_2O_3$ is known to be extremely sensitive to oxygen and to lose the characteristic $Ti^{3+}$–$Ti^{3+}$ dimer structures due to oxidation [23]. Figure 1(c) shows the Ti $2p$ core-level spectrum of the $Ti_2O_3$ films obtained at 100 K (insulating phase). Note that the existence of $Ti^{4+}$ states at 459 eV due to surface overoxidation [23] is hardly seen in the spectrum. The Ti $2p$ core level exhibits the complicated multiplet structure characteristic of $Ti_2O_3$ single crystals [8, 23]. The close similarities in the core-level spectra between the film surface and cleaved surface of $Ti_2O_3$ single crystals [8, 23] support the fact that the chemical states in the films are almost the same as those in bulk. From the cluster-model calculation, the satellite structures were attributed to the strong bonding of the $a_{1g}$ orbitals at the Ti–Ti dimer in $Ti_2O_9$ clusters [8]. Therefore, the existence of the multiplet structures demonstrates the formation of $Ti^{3+}$–$Ti^{3+}$ dimers even in the surface region of the $Ti_2O_3$ films.

The $Ti^{3+}$ ($3d^1$) states in the $Ti_2O_3$ films are also confirmed by the valence-band spectrum (Fig. 1(d)). The spectrum mainly consists of two features: a structure derived from O $2p$ states at binding energies of 4–10 eV and a two-peak structure derived from the Ti $3d$ states near $E_F$ [8, 20–23]. The Ti $3d$ states consist of a sharp coherent peak in the vicinity of $E_F$ and a weak broad satellite structure centered around 2.3 eV, demonstrating the $Ti^{3+}$ ($3d^1$) feature of $Ti_2O_3$. The valence bands initially seem to be almost the same between the films and bulk, which is consistent with the results of the Ti $2p$ core-level spectra, indicating that the properties of $Ti_2O_3$ are retained in the surface region of the films. However, a closer inspection of $E_F$ (inset of Fig. 1(d)) reveals that there is a small but distinct density of states at $E_F$ in the films. Furthermore, the coherent peak of the films is shifted toward the lower-binding-energy side from that of the bulk, reflecting the higher conductivity in the films (see Fig. S2 in the Supplemental Material [50]). Based on the shift of the peak positions, the chemical-potential shift due to excess hole carriers is evaluated to be approximately 200 meV in the $Ti_2O_3$ films. We will discuss the value of the chemical-potential shift in connection with the ARPES results.

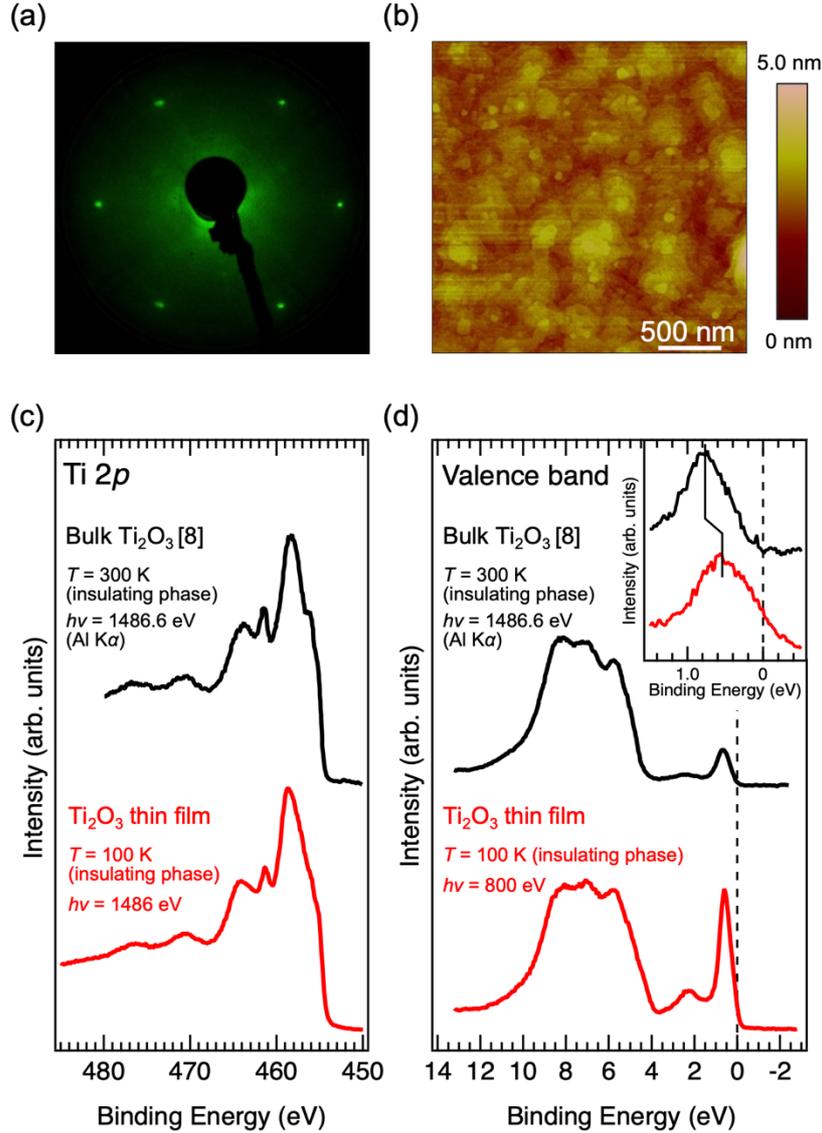

**FIG. 1.** (a) LEED pattern and (b) AFM image of the $Ti_2O_3$ films grown on $\alpha$-$Al_2O_3$ (0001) substrates. The hexagonal spots of the LEED pattern correspond to the surface BZ projected along the [111] direction of rhombohedral $Ti_2O_3$. The root-mean-square of surface roughness of the films was estimated to be less than 1.0 nm in the AFM image with a $10 \times 10$ μm$^2$ area, demonstrating the flat surface necessary for the present spectroscopic studies. (c) Ti 2$p$ core-level PES spectrum of the $Ti_2O_3$ films taken at 100 K (insulating phase) with a photon energy of $h\nu = 1486$ eV. (d) Corresponding valence-band spectrum taken at 800 eV. For comparison, the PES spectra of the cleaved surface of a $Ti_2O_3$ single crystal (black line) in an insulating phase are also shown for (c) and (d) [8]. The inset of (d) shows the comparison of the Ti 3$d$-derived coherent peaks near $E_F$. A clear peak shift of approximately 200 meV is observed, reflecting the hole-doped nature of the films, whereas the shape itself remains unchanged.

From the above characterizations, we have confirmed that the $Ti_2O_3$ films have a long-range ordered crystalline surface without the detectable overoxidation ($Ti^{4+}$ states) required for ARPES measurements and that characteristic $Ti^{3+}$–$Ti^{3+}$ dimers are formed in the surface region of the film accessible for SX-ARPES. Thus, we address the band structure of the $Ti_2O_3$ films by ARPES. Figure 2(a) shows the schematic crystal structure of $Ti_2O_3$ in both the primitive rhombohedral unit cell and associated conventional hexagonal cell. The corresponding rhombohedral BZ is depicted in Fig. 2(b), together with an equivalent hexagonally shaped one [10]. For notational simplicity, we refer to the high-symmetry points and lines in the hexagonal BZ hereafter.

Figure 2(c) shows the out-of-plane FS map for the $Ti_2O_3$ films obtained from normal-emission ARPES measurements for the Γ–A–H–K emission plane (the blue hatched area in the BZ shown in Fig. 2(b)). Reflecting the hole-doped nature, a meandering FS that follows the periodicity of the bulk BZ is clearly observed along the BZ center line (see also Fig. S12 in the Supplemental Material [50]). The cross-section of the FS is largest at the A point, monotonically decreases away from the A point, and is smallest at the Γ point. The observed FS topology was well reproduced by the DFT + $U$ calculation with reasonable parameters (Figs. S8 and S9 in the Supplemental Material [50]), which is discussed later. The in-plane FSs at the A–H–L emission plane (taken at 500 eV) and Γ–K–M emission plane (taken at 565 eV) are shown in Figs. 2(d). A triangular-like FS is clearly observed at the A point. The overall threefold intensity pattern of the observed FS is responsible for the trigonal symmetry of $Ti_2O_3$, suggesting the bulk origin of the FS. Indeed, the triangular-like shape of the FS was reversed at the other A point of one point below [41] (see Fig. S14 in the Supplemental Material [50]). At the Γ point, the FS with a round shape is also observed. These results indicate the existence of an open FS in the $Ti_2O_3$ films. It should be noted that we did not observe any indication of other FSs outside the open

FS in these measurement planes. Meanwhile, the existence of another small closed hole FS around the A point is predicted from band-structure calculations [10, 50]. However, it is difficult to examine the existence of another FS inside the open FS owing to the overlap of open FS having predominant intensity (see Fig. S9 in the Supplemental Material [50]), although the existence of the other FS is suggested by normal-emission ARPES along the $\Gamma$–A direction (see Fig. S12 in the Supplemental Material [50]). Assuming a triangular-pyramidal shape of the open FS, the carrier density of the $Ti_2O_3$ films would be approximately $2.4 \times 10^{20}$ cm$^{-3}$, which is consistent with the carrier density estimated from Hall effect measurements ($1.1 \times 10^{20}$ cm$^{-3}$) [16].

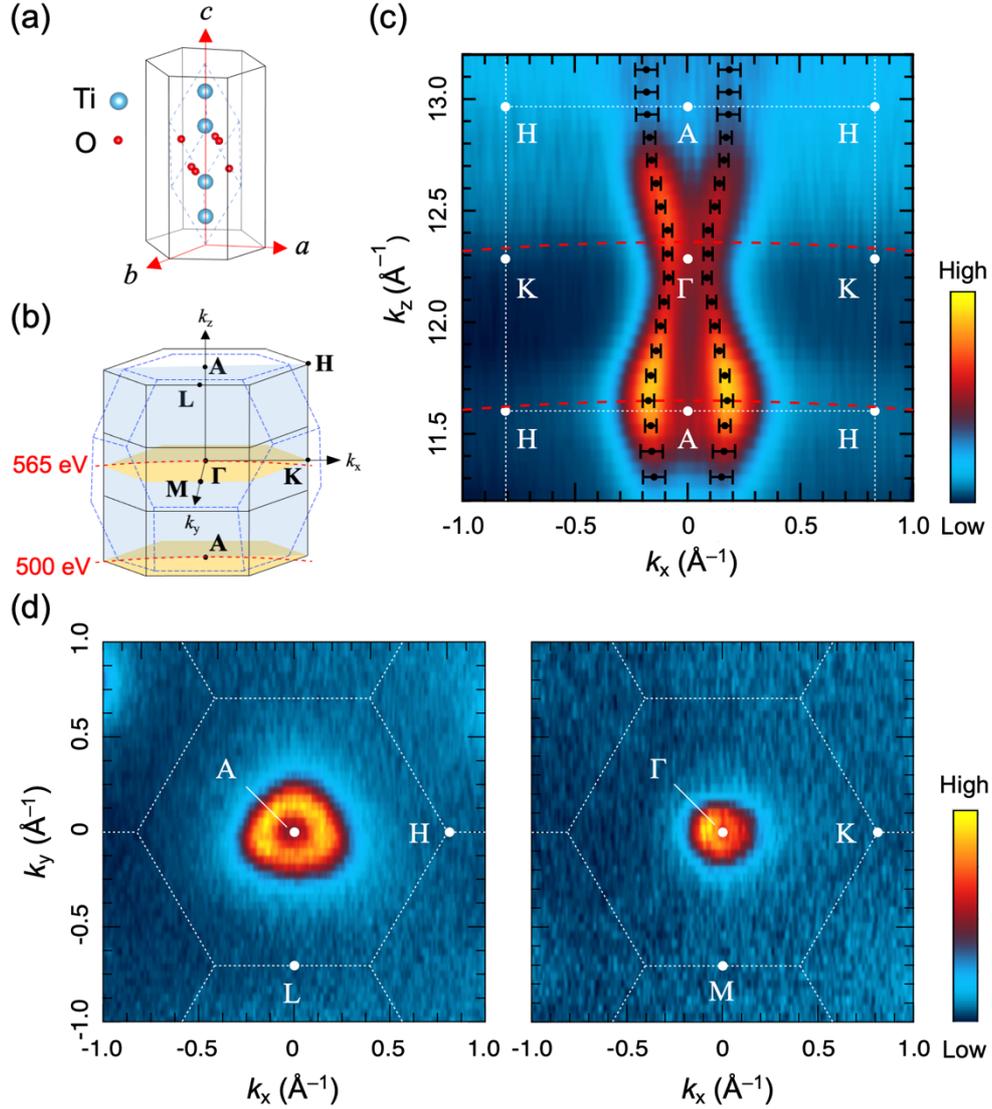

**FIG. 2.** (a) Crystal structure of $Ti_2O_3$ in primitive rhombohedral unit cell (blue dashed lines), surrounded by the conventional non-primitive hexagonal unit cell (black solid lines). The primitive rhombohedral unit cell contains two formula units, while the conventional hexagonal unit cell contains eighteen formula units. The red hexagonal unit vectors were determined to be $a = 5.102$ Å and $c = 13.80$ Å by XRD measurements (see Fig. S1 in the Supplemental Material [50]). (b) Corresponding rhombohedral BZ and an equivalent hexagonally-shaped one. In the hexagonal BZ, the high-symmetry points are labeled. The red dotted arc lines represent the **k** paths passing through the A point (at a photon energy of 500 eV) and the Γ point (at a photon energy of 565 eV). A light-blue plate represents the Γ–A–H–K emission plane where the out-of-plane FS map was measured, while two light-yellow plates represent the A–H–L and Γ–K–M emission planes for the in-plane FS mappings at corresponding energies. (c) Out-of-plane FS

map in the Γ–A–H–K emission plane obtained by varying the excitation photon energies from 470 to 640 eV. The Fermi-momentum ($k_F$) points determined by the ARPES spectra are indicated by data markers. (d) In-plane FS maps acquired at the constant photon energies of 500 eV (left panel: the A–H–L emission plane) and 565 eV (right panel: the Γ–K–M emission plane) by changing emission angles. The corresponding hexagonal BZ boundaries are overlaid as white lines. The FS maps were obtained by plotting the ARPES intensity within the energy window of $E_F \pm 50$ meV.

According to the band-structure calculation [10, 50], the electronic band structure near $E_F$ of $Ti_2O_3$ mainly consists of the $e_g^{\pi}$- and $a_{1g}$-derived bands. The former forms flat (dispersive) bands, while the latter forms dispersive (flat) bands along the out-of-plane (in-plane) direction, reflecting their anisotropic orbital shape. The $e_g^{\pi}$-derived band forms an electron pocket(s) at the Γ point, whereas the $a_{1g}$-derived band forms a hole pocket(s) at the A point. The slight overlap between the two pockets makes intrinsic $Ti_2O_3$ a compensated semimetal. This semimetallic nature may be the origin of the unusual physical properties of $Ti_2O_3$ because the delicate balance between the energy of the $e_g^{\pi}$ and $a_{1g}$ states causes a notable change in the conduction-carrier character as a result of hybridization with the other states. Thus, it is worth investigating the band structure near $E_F$. Figure 3 shows the experimental band structures along the in-plane high-symmetry directions (the Γ–K and A–H lines in Fig. 3(a) and 3(b), respectively). The corresponding energy distribution curves (EDCs) along each line are shown in Fig. S15 [50]. As expected from the FS maps in Fig. 2, highly dispersive hole bands exist both at the Γ and A points and form an FS at the zone center, reflecting the existence of hole carriers. The bands cross $E_F$ at Fermi momentum $k_F$ = 0.096 Å$^{-1}$ (0.155 Å$^{-1}$) around the Γ (A) point, while they disperse down to approximately 700 meV around the zone boundary.

To investigate the band structure near $E_F$ in more detail, we compared the experimental band structure with the DFT calculation based on the GGA + $U$ approximation using the reasonable parameters consistent with the experimental facts [17,50]. According to the DFT + $U$ calculation, the band structures, especially the FS topology, of $Ti_2O_3$ strongly depend on the parameters of $U$ and $c/a$ ratio (Figs. S5–S7 in the Supplemental Material [50]). Herein, we used $U = 2.2$ eV to reproduce the insulating phase of bulk $Ti_2O_3$, which exhibits an energy gap of approximately 100 meV at a $c/a$ ratio of 2.639 [17]. Moreover, the $c/a$ ratio, which also governs the electronic properties of $Ti_2O_3$, was fixed at 2.70 as determined from the XRD measurements (see Fig. S1 in the Supplemental Material [50]). The results of the DFT calculations are shown in the panels of Fig. 3, where the Fermi level in the DFT calculations is shifted by 180 meV toward the higher-binding-energy side to reproduce the $k_F$ value of the hole pocket at the $\Gamma$ point. This is because the electron pocket is always formed at the $\Gamma$ point in DFT calculations irrespective of $U$ values and $c/a$ ratios for $Ti_2O_3$ with metallic ground states (see Figs. S5–S7 in the Supplemental Material [50]). It should be noted that the shift of the Fermi level is consistent with the chemical-potential shift estimated from the shift of the Ti $3d$ coherent peak, as seen in the inset of Fig. 1(d).

The ARPES results and DFT calculations show fairly good agreement. According to the orbital projected band structures shown in Fig. 3 (also see Fig. S4 in the Supplemental Material [50]), the hole band centered at the $\Gamma$ point is assigned to the $e_g^\pi$-derived band. The hole band has a substantial $e_g^\pi$ character at the $\Gamma$ point, while it loses the $e_g^\pi$ character rapidly apart from the zone center owing to strong hybridization with other states. Meanwhile, the other hole FSs centered at the A point fundamentally consist of two $a_{1g}$-derived hole bands, which degenerate along the A–H line. The $a_{1g}$-derived hole bands have an $a_{1g}$ character at the A point, while they are strongly hybridized with the $e_g^\pi$ and $e_g^\sigma$ states apart from the zone center. As can be seen in Fig. 3, the present ARPES results predominately show the band to have significant $e_g^\pi$-orbital weight owing to the dipole selection rules for $p$ polarization of the incident light and measurement geometry used [33, 34, 57].

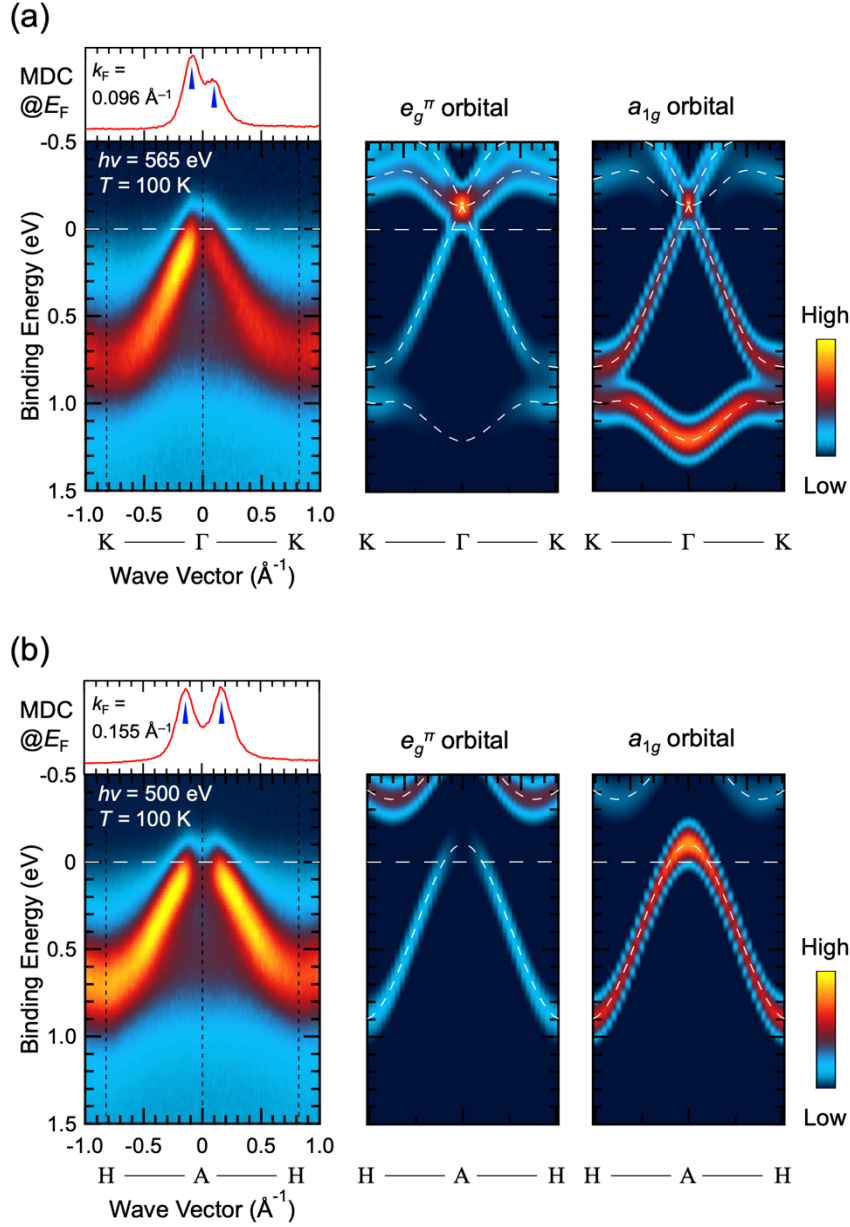

**FIG. 3.** Experimental band structure of the Ti$_2$O$_3$ films obtained by ARPES measurements (left panels) along the (a) Γ–K and (b) A–H high-symmetry lines, together with $e_g^\pi$ (middle panels) and $a_{1g}$ (right panels) orbital projected band structures. The corresponding **k** paths are illustrated in Fig. 2(c). Note that the experimental band structure is slightly asymmetric at the zone boundary, reflecting the rhombohedral BZ. The momentum distribution curves (MDCs) at $E_F$ with an energy window of ±5 meV are shown in the upper panel for each experimental band structure. The blue triangles indicate the Fermi momentum ($k_F$). In the orbital projections, the band structure is overlaid by white dotted lines.

Next, to demonstrate the importance of electron-electron correlation, we compare the band dispersion determined by ARPES with the DFT calculation in more detail. Figure 4 shows the band dispersion drawn based on the ARPES peak positions determined from the EDCs and MDCs, together with the DFT + $U$ calculations. The band-structure calculation was also conducted at $U = 0$ eV as a reference for clarifying the effect of the electron–electron correlation. Note that the band dispersion of $U = 0$ eV is almost identical to that previously reported [10]. As can be seen in Fig. 4, the overall band structure is in good agreement with the DFT calculation at $U = 2.2$ eV; the dispersion of the $e_g^{\pi}$-derived hole band centered at the $\Gamma$ point is quite well-reproduced. In addition, $k_F$ along the A–H direction shows excellent agreement between the experiment and calculation, indicating the validity of $U = 2.2$ eV and the Fermi level correction for describing the electronic structures of the hole-doped $Ti_2O_3$ films.

Meanwhile, in the calculation, a flat band exists at a higher binding energy along the $\Gamma$–K direction. The flat band originates from the $a_{1g}$ orbital and forms a hole pocket at the A point, reflecting the anisotropic feature of the $a_{1g}$ orbital with a large distribution along the [0001] direction (Fig. S4 in the Supplemental Material [50]). The band is barely seen in the ARPES images in Fig. 3(a), although the existence of the $a_{1g}$-derived band itself is confirmed by normal emission ARPES along the $\Gamma$–A direction (see Figs. S12 and S13 in the Supplemental Material [50]). The very weak intensity is probably caused by the dipole selection rules for the $a_{1g}$ orbital in the present experimental geometry [36–40], as demonstrated in Fig. 3(a). Furthermore, the presence of a broad nondispersive component at 0.5–0.6 eV masks the details of the band dispersion along the $\Gamma$–K direction. Therefore, it is difficult to determine the energy position of the band at the $\Gamma$ point from the present data. The nondispersive states can also be seen in the ARPES image along the A–H direction (Fig. 3(b)), where such a flat band is not predicted by the DFT calculation (Fig. 4(b)). It should be noted that in contrast to the case of $CrO_2$ [43], such nondispersive states cannot be found in the DFT calculation even with increasing $U$ (see Fig. S5 in the Supplemental Material [50]) for the case of $Ti_2O_3$. Although the origin of the nondispersive states is currently not clear, similar nondispersive quasi-localized states other than the lower Hubbard band (incoherent part) have also been observed in $V_2O_3$ [42], implying that these nondispersive states are a common feature at the surface of corundum-type conductive oxides.

In contrast to $U = 2.2$ eV, there is less agreement between the experiment and calculation at $U = 0$ eV. In particular, a significant discrepancy is observed in the $a_{1g}$-derived bands along the A–H direction (Fig. 4(b)). Note that we attempted to adjust the Fermi level for $U = 0$ eV to improve this discrepancy, but we did not find any improvement in the $k_F$ positions along with both the Γ–K and A–H directions. These results suggest the importance of the electron–electron correlation for describing the electronic structure of $Ti_2O_3$.

The disagreement between the experiment and calculation at $U = 0$ eV likely originates from the difference in the energy levels of the $a_{1g}$ and $e_g^\pi$ states. For the band dispersions in the DFT + $U$ calculations with varying $U$ (see Figs. S5 and S6 in the Supplemental Material [50]), the energy separation between the $a_{1g}$ and $e_g^\pi$ states increases with increasing $U$; the $a_{1g}$-derived band is pushed down, whereas the $e_g^\pi$-derived band is pushed up. The increment of $U$ causes narrowing of the $a_{1g}$-derived band dispersion along the A–H direction and widening of the $e_g^\pi$-derived band dispersion along the Γ–K direction. As a result, the electron and hole pockets at the Γ and A points, respectively, simultaneously become smaller, and eventually a tiny energy gap opens at $U = 2.5$ eV. Accordingly, the occupancy of the lowest-lying $a_{1g}$ state increases with increasing $U$.

The semimetallic ground states of $Ti_2O_3$ predominantly originate from the slight overlap between the $e_g^\pi$-derived electron pocket at the Γ point and the $a_{1g}$-derived hole pockets at the A point (at the midpoint of the Γ–A line for $U = 2.2$ eV) (see Fig. S5 in the Supplemental Material [50]). The increment of $U$ reduces the degree of the overlap. Therefore, the excellent agreements in $k_{FS}$ between the ARPES and DFT calculations at $U = 2.2$ eV indicate that the electron–electron correlation plays an essential role in the band structures of $Ti_2O_3$. The sensitivity of $k_F$ (FS topology) to $U$ triggers an additional check for the validity of $U = 2.2$ eV to describe the electronic band structure of the $Ti_2O_3$ films. Difference in $k_F$ at the A point between the ARPES and DFT calculations with varying $U$ (Figs. S10 and S11 in the Supplemental Material [50]) suggests that the best match is obtained at $U = 2.2$ eV. This indicates the validity of the $U$ value of 2.2 eV.

The importance of electron–electron correlation is further supported by the ARPES results shown in Fig. 4, where an almost identical bandwidth between the experiment and calculation is observed along both the Γ–K and A–H lines. Meanwhile, reduction of the overlapping was also achieved by shortening the Ti–Ti bond distances along the $c$-axis (or equivalently, by reducing the $c/a$ ratio) (see Fig. S7 in the Supplemental Material [50]), although an unusually short Ti–Ti distance of less than 2.2 Å was required to form the insulating ground state in $Ti_2O_3$ without $U$ [10]. Thus, it should be considered that the FS topology of $Ti_2O_3$ is governed by a delicate balance between the electron–electron correlations and lattice deformations and, consequently, is very sensitive to both. Besides, it should be bear in mind that the present study has not ruled out the possible contribution of dynamic electronic correlations to the MIT, namely the possibility of Ti–Ti dimer deformation assisted Mott transition in $Ti_2O_3$ [25]. Thus, a more elaborate band-structure calculation that adjusts both the effects and incorporates the dynamic electronic correlations in a realistic manner is necessary to reproduce the ARPES results as well as the physical properties of $Ti_2O_3$ [11, 24, 25].

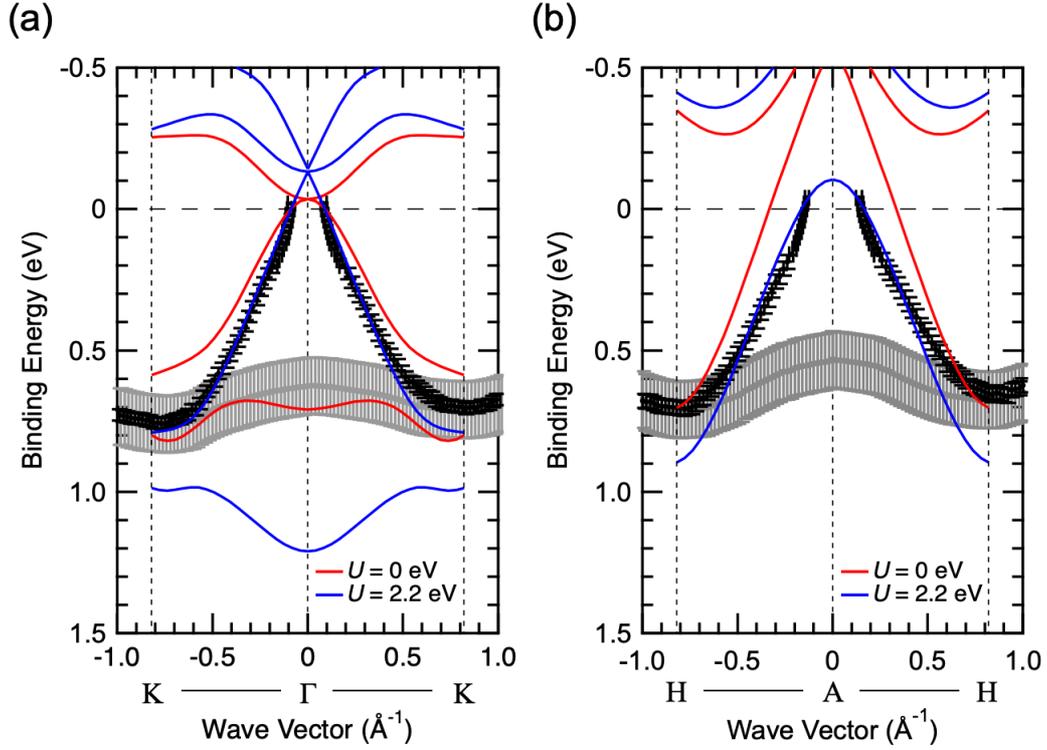

**FIG. 4.** Comparison of the experimental band structure and GGA+$U$ calculation. Peak positions determined from EDCs and MDCs are plotted by the data markers. The black and gray markers indicate the peak positions of the dispersive and nondispersive quasi-localized states, respectively. The results of GGA+$U$ calculation with $U = 0$ and 2.2 eV are plotted as the red and blue lines, respectively. The Fermi level of the band structure calculations with $U = 2.2$ eV (0 eV) was shifted by 180 meV (150 meV) toward the higher-binding-energy side to reproduce the experimental $k_F$ value at the $\Gamma$ point. The sudden upturn of band dispersion starting from $E_F$ may originate from the interplay of the energy resolution and Fermi-edge cutoff [58].

Finally, we discuss the possible origin of the MIT in $Ti_2O_3$. From the band structure shown in Fig. 4 (also see Fig. S3 in the Supplemental Material [50]), "intrinsic" $Ti_2O_3$ exhibits a semimetallic band structure, where the electron pocket at the $\Gamma$ point and hole pocket at the midpoint of the $\Gamma$–A line for $U = 2.2$ eV (at the A point for $U = 0$ eV) slightly overlap in energy. In general, for such a small carrier system, the influence of the electron–electron correlation on phase-transition phenomena is not significantly large [32], whereas a slight change in a band structure significantly changes the FS topology and resultant transport properties. Meanwhile, it is known that lattice deformation occurs in $Ti_2O_3$ regardless of the doping concentration [3, 7, 31]. Therefore, the most plausible scenario is that the MIT of $Ti_2O_3$ is a semimetal–semimetal or semimetal–semiconductor transition caused by the changes in the FS topology due to lattice deformation, not a filling-controlled MIT [32]. In fact, the "MIT temperature" of $(Ti_{1-x}V_x)_2O_3$ remains almost unchanged with $x$, even though the carrier density changes notably; consequently, the ground state changes from insulator at $x = 0$ to metal at $x = 0.06$ [7]. To verify the possibility of semimetal–semimetal or semimetal–semiconductor transition in $Ti_2O_3$ caused by the changes in the FS topology due to lattice deformation, further investigations are required; in particular, detailed temperature-dependent studies on both the crystal and electronic structures are necessary.

**CONCLUSION**

To investigate the electronic band structure near $E_F$ of $Ti_2O_3$, we performed SX-ARPES on single-crystalline epitaxial $Ti_2O_3$ films grown on $\alpha$-$Al_2O_3$ substrates. Using well-defined surfaces of the epitaxial films, we clearly observed the electronic band structure of $Ti_2O_3$. The Ti $3d$-derived band forms an open hole FS with a triangular-pyramidal vase shape along the $\Gamma$–A line in the hexagonal BZ, which is in line with the previous transport measurements. We also found the highly dispersive $e_g^\pi$- and $a_{1g}$-derived bands centered at the $\Gamma$ and A points, respectively. The observed band structures were compared with the band-structure calculations based on DFT with GGA + $U$ correction. Detailed analysis with varying $U$ revealed that the obtained band structure is well-described by the DFT calculation at $U = 2.2$ eV. These results suggest that the electron–electron correlation plays an important role in describing the overall electronic structure of $Ti_2O_3$. Nevertheless, the influence of electron–electron correlation on the quantum phase transition in $Ti_2O_3$ is considered to be weak owing to its low carrier density. Therefore, the ARPES and DFT calculation results presented here have important implications regarding the origin of the characteristic MIT in $Ti_2O_3$; the MIT is a semimetal–semimetal or semimetal–semiconductor transition caused by changes in the FS topology due to lattice deformation.

**ACKNOWLEDGMENTS**

The authors acknowledge R. Tokunaga, X. Cheng, and D. K. Nguyen for their support in the experiments at KEK-PF. The authors also thank R. Yukawa for sharing his program for analyzing ARPES data with us and for his assistance with data analysis. This work was financially supported by a Grant-in-Aid for Scientific Research (Nos. 16H02115, 16KK0107, 20KK0117, 22H01948, and 22H01947) from the Japan Society for the Promotion of Science (JSPS), CREST (JPMJCR18T1) from the Japan Science and Technology Agency (JST), the MEXT Element Strategy Initiative to Form Core Research Center (JPMXP0112101001), Asahi Glass Foundation, Iketani Science and Technology Foundation, Nagamori Foundation, Research Foundation for the Electrotechnology of Chubu, and Toyota Riken Scholar Program. N.H. acknowledges financial support from the Chemistry Personnel Cultivation Program of the Japan Chemical Industry Association. T.K. acknowledges financial support from the Division for Interdisciplinary Advanced Research and Education at Tohoku University. The work performed at KEK-PF was approved by the Program Advisory Committee (proposals 2019T004, 2018S2-004, 2021G621, and 2021S2-002) at the Institute of Materials Structure Science, KEK. The authors would also like to thank Editage (www.editage.com) for English language editing.

**References**

1. J. M. Honig and T. B. Reed, "Electrical Properties of $Ti_2O_3$ Single Crystals" Phys. Rev. **174**, 1020 (1968). DOI: 10.1103/PhysRev.174.1020.

2. J. M. Honig, "Nature of the Electrical Transition in $Ti_2O_3$" Rev. Mod. Phys. **40**, 748 (1968). DOI: 10.1103/RevModPhys.40.748.

3. J. J. Capponi, M. Marezio, J. Dumas, and C. Schlenker, "Lattice parameters variation with temperature of $Ti_2O_3$ and $(Ti_{0.98}V_{0.02})_2O_3$ from single crystal X-Ray data" Solid State Commun. **20**, 893 (1976). DOI: 10.1016/0038-1098(76)91299-0.

4. C. N. R. Rao, R. E. Loehman, and J. M. Honig, "Crystallographic study of the transition in $Ti_2O_3$" Phys. Lett. **27A**, 271 (1968). DOI: 10.1016/0375-9601(68)90696-8.

5. L. L. Van Zandt, J. M. Honig, and J. B. Goodenough, "Resistivity and magnetic order in $Ti_2O_3$" J. Appl. Phys. **39**, 594–595 (1968). DOI: 10.1063/1.2163536.

6. H. J. Zeiger, "Unified model of the insulator-metal transition in $Ti_2O_3$ and the high-temperature Transitions in $V_2O_3$" Phys. Rev. B **11**, 5132 (1975). DOI: 10.1103/PhysRevB.11.5132.

7. M. Uchida, J. Fujioka, Y. Onose, and Y. Tokura, "Charge dynamics in thermally and doping induced insulator-metal transitions of $(Ti_{1−x}V_x)_2O_3$" Phys. Rev. Lett. **101**, 066406 (2008). DOI: 10.1103/PhysRevLett.101.066406.

8. C. F. Chang, T. C. Koethe, Z. Hu, J. Weinen, S. Agrestini, L. Zhao, J. Gegner, H. Ott, G. Panaccione, H. Wu, M. W. Haverkort, H. Roth, A. C. Komarek, F. Offi, G. Monaco, Y. –F. Liao, K. –D. Tsuei, H. –J. Lin, C. T. Chen, A. Tanaka, and L. H. Tjeng, "*c*-Axis Dimer and Its Electronic Breakup: The Insulator-to-Metal Transition in $Ti_2O_3$" Phys. Rev. X **8**, 021004 (2018). DOI: 10.1103/PhysRevX.8.021004.

9. G. V. Chandrashekhar, Q. Won Choi, J. Moyo, and J. M. Honig, "The electrical transition in

V-doped Ti$_2$O$_3$" Mater. Res. Bull. 5,999–1007(1970). DOI: 10.1016/0025-5408(70)90048-6.

10. L. F. Mattheiss, "Electronic structure of rhombohedral Ti$_2$O$_3$" J. Phys.: Condens. Mater **8**, 5987 (1996). DOI: 10.1088/0953-8984/8/33/007.

11. V. Singh and J. J. Pulikkotil, "Electronic phase transition and transport properties of Ti$_2$O$_3$" J. Alloys Compd. **658**, 430 (2016). DOI: 10.1016/j.jallcom.2015.10.203.

12. Y. Tsujimoto, Y. Matsushita, S. Yu, K. Yamaura, T. Uchikoshi, "Size dependence of structural, magnetic, and electrical properties in corundum-type Ti$_2$O$_3$ nanoparticles showing insulator–metal transition" J. Asian Ceram. Soc. 3, 325–333 (2015). DOI: 10.1016/j.jascer.2015.06.007.

13. F. J. Morin, "Oxides which show a metal-to-insulator transition at the Neel temperature" Phys. Rev. Lett. **3**, 34 (1959). DOI:10.1103/PhysRevLett.3.34.

14. A. Tanaka, "On the metal-insulator transitions in VO$_2$ and Ti$_2$O$_3$ from a unified viewpoint" J. Phys. Soc. Jpn. **73**, 152 (2004). DOI: 10.1143/JPSJ.73.152.

15. H. Sato, A. Tanaka, M. Sawada, F. Iga, K. Tsuji, M. Tsubota, M. Takemura, K. Yaji, M. Nagira, A. Kimura, T. Takabatake, H. Namatame, and M. Taniguchi, "Ti 3$d$ orbital change across metal-insulator transition in Ti$_2$O$_3$: polarization-dependent soft x-ray absorption spectroscopy at Ti 2$p$ edge" J. Phys. Soc. Jpn. **75**, 053702 (2006). DOI: 10.1143/JPSJ.75.053702.

16. K. Yoshimatsu, H. Kurokawa, K. Horiba, H. Kumigashira, and A. Ohtomo, "Large anisotropy in conductivity of Ti$_2$O$_3$ films" APL Mater. **6**, 101101 (2018). DOI: 10.1063/1.5050823.

17. K. Yoshimatsu, N. Hasegawa, Y. Nambu, Y. Ishii, Y. Wakabayashi, and H. Kumigashira, "Metallic ground states of undoped Ti$_2$O$_3$ films induced by elongated $c$-axis lattice constant" Sci. Rep. **10**, 22109 (2020). DOI: 10.1038/s41598-020-79182-5.

18. Y. Guo, S. J. Clark, and J. Robertson, "Electronic and magnetic properties of Ti$_2$O$_3$, Cr$_2$O$_3$,

and $Fe_2O_3$ calculated by the screened exchange hybrid density functional" J. Phys.: Condens. Mater **24**, 325504 (2012). <u>DOI: 10.1088/0953-8984/24/32/325504</u>.

19. K. E. Smith and V. E. Henrich, "Bulk band dispersion in $Ti_2O_3$ and $V_2O_3$" Phys. Rev. B **38**, 5965 (1988). DOI: <u>10.1103/PhysRevB.38.5965</u>.

20. T. Uozumi, K. Okada, A. Kotani, Y. Tezuka, and S. Shin, "Ti $2p$ and resonant $3d$ photoemission spectra of $Ti_2O_3$" J. Phys. Soc. Jpn. **65**, 1150 (1996). DOI: <u>10.1143/jpsj.65.1150</u>.

21. J. M. McKay, M. H. Mohamed, and V. E. Henrich, "Localized $3p$ excitations in $3d$ transition-metal-series spectroscopy" Phys. Rev. B **35**, 4304 (1987). DOI: <u>10.1103/PhysRevB.35.4304</u>.

22. A. Mooradian and P. M. Raccah, "Raman study of the semiconductor-metal transition in $Ti_2O_3$" Phys. Rev. B 3, 4253 (1971). DOI: <u>10.1103/PhysRevB.3.4253</u>.

23. S. A. Chambers, M. H. Engelhard, L. Wang, T. C. Droubay, M. E. Bowden, M. J. Wahila, N. F. Quackenbush, L. F. J. Piper, Tien-Lin Lee, C. J. Nelin, and P. S. Bagus, "X-ray photoelectron spectra for single-crystal $Ti_2O_3$: Experiment and theory" Phys. Rev. B 96, 205143 (2017). DOI: <u>10.1103/PhysRevB.96.205143</u>.

24. F. Iori, M. Gatti, and A. Rubio, "Role of nonlocal exchange in the electronic structure of correlated oxides" Phys. Rev. B **85**, 115129 (2012). DOI: <u>10.1103/PhysRevB.85.115129</u>.

25. A. I. Poteryaev, A. I. Lichtenstein, and G. Kotliar, "Nonlocal Coulomb interaction and metal-insulator transition in $Ti_2O_3$: A cluster LDA + DMFT approach" Phys. Rev. Lett. **93**, 086401 (2004). DOI: <u>10.1103/PhysRevLett.93.086401</u>.

26. H. Nakatsugawa and E. Iguchi, "Transition phenomenon in $Ti_2O_3$ using the discrete variational $X\alpha$ cluster method and periodic shell model" Phys. Rev. B **56**, 12931 (1997). DOI: <u>10.1103/PhysRevB.56.12931</u>.

27. G. V. Chandrashekhar, Q. Won Choi, J. Moyo, and J. M. Honig, "The electrical transition in V-doped $Ti_2O_3$" Mat. Res. Bull. **5**, 999 (1970). DOI: 10.1016/0025-5408(70)90048-6.

28. S. H. Shin, R. L. Aggarwal, B. Lax, and J. M. Honig, "Raman scattering in $Ti_2O_3$-$V_2O_3$ alloys" Phys. Rev. B 9, 583 (1974). DOI: 10.1103/PhysRevB.9.583.

29. S. H. Shin, F. H. Pollak, T. Halpern, P. M. Raccah, "Resonance Raman scattering in $Ti_2O_3$ in the range 1.8–2.7 eV" Solid State Commun. 16, 687 (1975). DOI: 10.1016/0038-1098(75)90453-6.

30. A. Tanaka, "A New Scenario on the Metal–Insulator Transition in $VO_2$" J. Phys. Soc. Jpn. 72, 2433–2436 (2003). DOI: 10.1143/JPSJ.72.2433.

31. C. E. Rice and W. R. Robinson "Structural changes in the solid solution $(Ti_{1-x}V_x)_2O_3$ as a varies from zero to one" J. Solid State Chem. **21**, 145 (1977). DOI: 10.1016/0022-4596(77)90154-2

32. M. Imada, A. Fujimori, and Y. Tokura, "Metal-insulator transitions" Rev. Mod. Phys. 70, 1039 (1998). DOI: 10.1103/RevModPhys.70.1039.

33. A. Damascelli, "Probing the Electronic Structure of Complex Systems by ARPES" Phys. Scr. 61, T109 (2004). DOI: 10.1238/Physica.Topical.109a00061.

34. A. Damascelli, Z. Hussain, and Z.-X. Shen, "Angle-resolved photoemission studies of the cuprate superconductors" Rev. Mod. Phys. **75**, 473 (2003). DOI: 10.1103/RevModPhys.75.473.

35. J. A. Sobota, Y. He, and Z.-X. Shen, "Angle-resolved photoemission studies of quantum materials" Rev. Mod. Phys. **93**, 025006 (2021). DOI: 10.1103/RevModPhys.93.025006.

36. M. Kobayashi, K. Yoshimatsu, E. Sakai, M. Kitamura, K. Horiba, A. Fujimori, and H. Kumigashira, "Origin of the Anomalous Mass Renormalization in Metallic Quantum Well States of Strongly Correlated Oxide $SrVO_3$" Phys. Rev. Lett. **115**, 076801 (2015). DOI:

10.1103/PhysRevLett.115.076801.

37. M. Kobayashi, K. Yoshimatsu, T. Mitsuhashi, M. Kitamura, E. Sakai, R. Yukawa, M. Minohara, A. Fujimori, K. Horiba, and H. Kumigashira, "Emergence of Quantum Critical Behavior in Metallic Quantum-Well States of Strongly Correlated Oxides" Sci. Rep. **7**, 16621 (2017). DOI: 10.1038/s41598-017-16666-x.

38. T. Mitsuhashi, M. Minohara, R. Yukawa, M. Kitamura, K. Horiba, M. Kobayashi, and H. Kumigashira, "Influence of $k_\perp$ broadening on ARPES spectra of the (110) and (001) surfaces of SrVO$_3$ films" Phys. Rev. B **94**, 125148 (2016). DOI: 10.1103/PhysRevB.94.125148.

39. T. Kanda, D. Shiga, R. Yukawa, N. Hasegawa, D.K. Nguyen, X. Cheng, R. Tokunaga, M. Kitamura, K. Horiba, K. Yoshimatsu, and H. Kumigashira, "Electronic structure of SrTi$_{1-x}$V$_x$O$_3$ films studied by in situ photoemission spectroscopy: Screening for a transparent electrode material" Phys. Rev. B **104**, 115121 (2021). DOI: 10.1103/PhysRevB.104.115121.

40. R. Yukawa, M. Kobayashi, T. Kanda, D. Shiga, K. Yoshimatsu, S. Ishibashi, M. Minohara, M. Kitamura, K. Horiba, A. F. Santander-Syro, and H. Kumigashira, "Resonant tunneling driven metal-insulator transition in double quantum-well structures of strongly correlated oxide" Nat. Commun. **12**, 7070 (2021). DOI: 10.1038/s41467-021-27327-z.

41. I. Lo Vecchio, J. D. Denlinger, O. Krupin, B. J. Kim, P. A. Metcalf, S. Lupi, J. W. Allen, and A. Lanzara, "Fermi surface of metallic V$_2$O$_3$ from angle-resolved photoemission: mid-level filling of $e_g^\pi$ bands" Phys. Rev. Lett. **117**, 166401 (2016). DOI: 10.1103/PhysRevLett.117.166401.

42. M. Thees, M. –H. Lee, R. L. Bouwmeester, P. H. Rezende-Gonçalves, E. David, A. Zimmers, F. Fortuna, E. Frantzeskakis, N. M. Vargas, Y. Kalcheim, P. Le Fèvre, K. Horiba, H. Kumigashira, S. Biermann, J. Trastoy, M. J. Rozenberg, I. K. Schuller, and A. F. Santander-Syro, "Imaging the itinerant-to-localized transmutation of electrons across the metal-to-insulator transition in $V_2O_3$" Sci. Adv. **7**, eabj1164 (2021). DOI: <u>10.1126/sciadv.abj1164</u>.

43. F. Bisti, V. A. Rogalev, M. Karolak, S. Paul, A. Gupta, T. Schmitt, G. Güntherodt, V. Eyert, G. Sangiovanni, G. Profeta, and V. N. Strocov, "Weakly-correlated nature of ferromagnetism in nonsymmorphic $CrO_2$ revealed by bulk-sensitive soft-x-ray ARPES" Phys. Rev. X **7**, 041067 (2017). DOI: <u>10.1103/PhysRevX.7.041067.</u>

44. V. N. Strocov, A. Chikina, M. Caputo, M. –A. Husanu, F. Bisti, D. Bracher, T. Schmitt, F. Miletto Granozio, C. A. F. Vaz, and F. Lechermann, "Electronic phase separation at $LaAlO_3$/$SrTiO_3$ interfaces tunable by oxygen deficiency" Phys. Rev. Mater. **3**, 106001 (2019). DOI: <u>10.1103/PhysRevMaterials.3.106001.</u>

45. A. Chikina, D. V. Christensen, V. Borisov, M. –A. Husanu, Y. Chen, X. Wang, T. Schmitt, M. Radovic, N. Nagaosa, A. S. Mishchenko, R. Valentí, and V. N. Strocov, "Band-order anomaly at the $\gamma$-$Al_2O_3$/$SrTiO_3$ interface drives the electron-mobility boost" ACS Nano **15**, 4347–4356 (2021). DOI: <u>10.1021/acsnano.0c07609.</u>

46. V. N. Strocov, L. L. Ley, M. Kobayashi, C. Cancellieri, M. –A. Husanu, A. Chikina, N. B. M. Schröter, X. Wang, J. A. Krieger, and Z. Salman, "k-resolved electronic structure of buried heterostructure and impurity systems by soft-X-ray ARPES" J. Electron Spectrosc. Relat. Phenom. **236**, 1–8 (2019). DOI: <u>10.1016/j.elspec.2019.06.009.</u>

47. L. L. Lev, J. Krempaský, U. Staub, V. A. Rogalev, T. Schmitt, M. Shi, P. Blaha, A. S. Mishchenko, A. A. Veligzhanin, Y. V. Zubavichus, M. B. Tsetlin, H. Volfová, J. Braun, J. Minár, and V. N. Strocov, "Fermi surface of three-dimensional $La_{1-x}Sr_xMnO_3$ explored by soft-x-ray ARPES: Rhombohedral lattice distortion and its effect on magnetoresistance" Phys. Rev. Lett. **114**, 237601 (2015). DOI: <u>10.1103/PhysRevLett.114.237601.</u>

48. M. –A. Husanu, L. Vistoli, C. Verdi, A. Sander, V. Garcia, J. Rault, F. Bisti, L. L. Lev, T. Schmitt, F. Giustino, A. S. Mishchenko, M. Bibes, and V. N. Strocov, "Electron-polaron dichotomy of charge carriers in perovskite oxides" Commun. Phys. **3**, 62 (2020). DOI: 10.1038/s42005-020-0330-6.

49. G. Berner, M. Sing, H. Fujiwara, A. Yasui, Y. Saitoh, A. Yamasaki, Y. Nishitani, A. Sekiyama, N. Pavlenko, T. Kopp, C. Richter, J. Mannhart, S. Suga, and R. Claessen, "Direct *k*-sapce mapping of the electronic structure in an oxide-oxide interface" Phys. Rev. Lett. **110**, 247601 (2013). DOI: 10.1103/PhysRevLett.110.247601.

50. See Supplemental Material at http://link.aps.org/supplemental/X for preparation of $TiO_x$ ceramic target, crystal structures of the films, transport properties of the films, band-structure calculations, and analysis details of the ARPES data.

51. K. Yoshimatsu, O. Sakata, and A. Ohtomo, "Superconductivity in $Ti_4O_7$ and $\gamma$-$Ti_3O_5$ films" Sci. Rep. **7**, 12544 (2017). DOI: 10.1038/s41598-017-12815-4.

52. H. Kurokawa, K. Yoshimatsu, O. Sakata, and A. Ohtomo, "Effects of phase fraction on superconductivity of low-valence eutectic titanate films" J. Appl. Phys. **122**, 055302 (2017). DOI: 10.1063/1.4997443.

53. P. Giannozzi, S. Baroni, N. Bonini, M. Calandra, R. Car, C. Cavazzoni, D. Ceresoli, G. L. Chiarotti, M. Cococcioni, I. Dabo, A. Dal Corso, S. de Gironcoli, S. Fabris, G. Fratesi, R. Gebauer, U. Gerstmann, C. Gougoussis, A. Kokalj, M. Lazzeri, L. Martin-Samos, N. Marzari, F. Mauri, R. Mazzarello, S. Paolini, A. Pasquarello, L. Paulatto, C. Sbraccia, S. Scandolo, G. Sclauzero, A. P. Seitsonen, A. Smogunov, P. Umari and R. M. Wentzcovitch, "QUANTUM ESPRESSO: a modular and open-source software project for quantum simulations of materials" J. Phys.: Condens. Matter **21**, 395502 (2009). DOI: 10.1088/0953-8984/21/39/395502.

54. P. Giannozzi, O. Andreussi, T. Brumme, O. Bunau, M. B. Nardelli, M. Calandra, R. Car, C. Cavazzoni, D. Ceresoli, M. Cococcioni, N. Colonna, I. Carnimeo, A. Dal Corso, S. de

Gironcoli, P. Delugas, R. A. DiStasio, Jr., A. Ferretti, A. Floris, G. Fratesi, G. Fugallo, R. Gebauer, U. Gerstmann, F. Giustino, T. Gorni, J. Jia, M. Kawamura, H.-Y. Ko, A. Kokalj, E. Küçükbenli, M. Lazzeri, M. Marsili, N. Marzari, F. Mauri, N. L. Nguyen, H.-V. Nguyen, A. Otero-de-laRoza, L. Paulatto, S. Poncé, D. Rocca, R. Sabatini, B. Santra, M. Schlipf, A. P. Seitsonen, A. Smogunov, I. Timrov, T. Thonhauser, P. Umari, N. Vast, X. Wu and S. Baroni, "Advanced capabilities for materials modelling with Quantum ESPRESSO" J. Phys.: Condens. Matter **29**, 465901 (2017). DOI: 10.1088/1361-648X/aa8f79.

55. J. P. Perdew, K. Burke, and M. Ernzerhof, "Generalized gradient approximation made simple" Phys. Rev. Lett. **77**, 3865 (1996). DOI: 10.1103/PhysRevLett.77.3865.

56. H. J. Monkhorst and J. D. Pack, "Special points for Brillouin-zone integrations" Phys. Rev. B **13**, 5188 (1976). DOI: 10.1103/PhysRevB.13.5188.

57. S. Hüfner, "Photoelectron Spectroscopy" (Springer-Verlag, Berlin,1995). ISBN: 978-3-662-03150-6

58. G. Levy, W. Nettke, B. M. Ludbrook, C. N. Veenstra, and A. Damascelli, "Deconstruction of resolution effects in angle-resolved photoemission" Phys. Rev. B **90**, 045150 (2014). DOI: 10.1103/PhysRevB.90.045150.

# Supplemental Material

## Electronic band structure of Ti$_2$O$_3$ thin films studied by angle-resolved photoemission spectroscopy

Naoto Hasegawa[1], Kohei Yoshimatsu[1,2,*], Daisuke Shiga[1], Tatsuhiko Kanda[1], Satoru Miyazaki[1], Miho Kitamura[3], Koji Horiba[3], and Hiroshi Kumigashira[1,2,3]

[1] *Institute of Multidisciplinary Research for Advanced Materials (IMRAM), Tohoku University, Sendai, 980–8577, Japan*

[2] *Materials Research Center for Element Strategy (MCES), Tokyo Institute of Technology, Yokohama 226–8503, Japan*

[3] *Photon Factory, Institute of Materials Structure Science, High Energy Accelerator Research Organization (KEK), Tsukuba, 305–0801, Japan*

*Correspondence:   kohei.yoshimatsu.c6@tohoku.ac.jp

# 1. Sample growth and characterization

## 1.1 Preparation of Ti₂O₃ ceramic target

A ceramic target for laser ablation with a nominal composition of $Ti_2O_3$ was prepared using a solid-state reaction method. Ti (3N purity) and $TiO_2$ (4N purity) powders with a molar ratio of 1:3 were mixed and then pelletized. The pellet was sintered at 1100°C for 12 h with continuous Ar/H₂ gas flow (0.5 L/min) to achieve a highly reductive atmosphere in which the $Ti_2O_3$ phase could be stabilized.

## 1.2 Crystal structure of Ti₂O₃ films

Figure S1(a) shows the out-of-plane X-ray diffraction (XRD) pattern of the $Ti_2O_3$ films on $\alpha$-$Al_2O_3$ (0001) substrates. Two film-derived peaks were detected at $2\theta$ of approximately 39.12° and 84.10°, corresponding to the $Ti_2O_3$ 0006 and 000$\underline{12}$ reflections, respectively. From the reflections, the $c$-axis lattice constant was estimated to be 13.80 Å. The full width at half maximum of the $Ti_2O_3$ 0006 reflection is 0.057° in the $\omega$-scan rocking curve profile (inset of Fig. S1(a)), confirming the high crystallinity of the obtained films.

The $a$-axis lattice constant of the $Ti_2O_3$ films was determined from the reciprocal space map shown in Fig. S1(b). The $a$-axis lattice constant was estimated to be 5.102 Å from the $Ti_2O_3$ 10-1$\underline{10}$ reciprocal point, and thus, the $c/a$ ratio is 2.70. The $c/a$ ratio is larger than that of $Ti_2O_3$ bulk single crystals ($c/a$ = 2.639 from $a$ = 5.157 Å and $c$ = 13.61 Å) [3]. Note that the obtained lattice constants are almost identical to those in previous reports [16,17], where the films were grown under almost the same conditions. In the present ARPES analyses and density functional theory (DFT) calculations, we adopted the experimentally determined lattice parameters ($a$ = 5.102 Å, $c$ = 13.80 Å, and $c/a$ = 2.70).

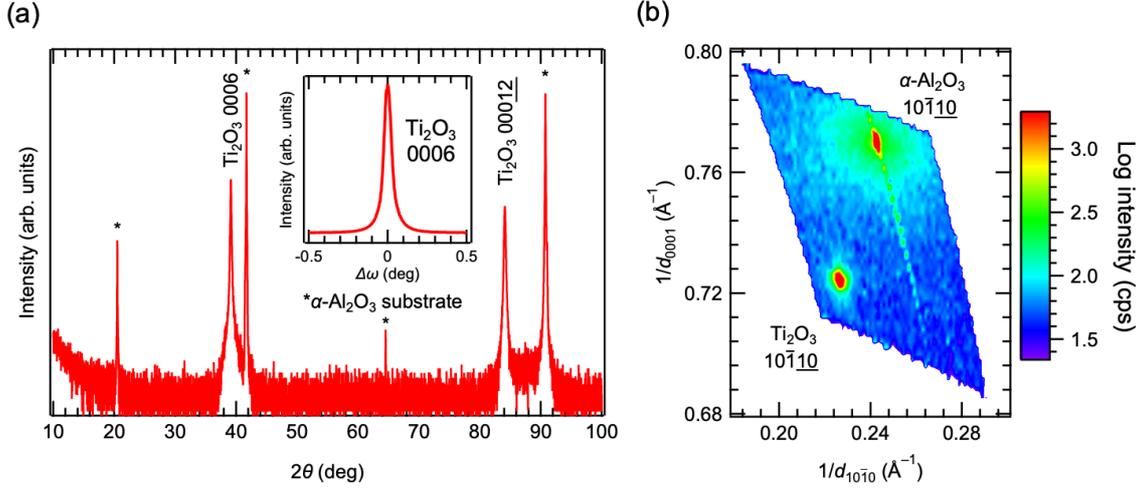

**FIG. S1.** (a) Out-of-plane XRD pattern of the $Ti_2O_3$ films grown on $\alpha$-$Al_2O_3$ (0001) substrates. The inset shows the $\omega$-scan rocking curve profile of the $Ti_2O_3$ 0006 reflection. (b) Reciprocal space map of the $Ti_2O_3$ films around the $\alpha$-$Al_2O_3$ 10-1$\underline{10}$ reciprocal point.

### 1.3 Transport properties of $Ti_2O_3$ films

Figure S2 shows the temperature dependence of the resistivity ($\rho$–$T$) for the $Ti_2O_3$ films. The $\rho$–$T$ curves of the $Ti_2O_3$ bulk single crystals [7] and previously reported $Ti_2O_3$ films [16] are also shown for comparison. The present $Ti_2O_3$ films showed a broad metal–insulator transition (MIT) at approximately 250 K, accompanied by an order of magnitude change in resistivity. The characteristic $\rho$–$T$ curve around the MIT temperature ($T_{MIT}$) is in good agreement with that of previously reported $Ti_2O_3$ films with a hole concentration of $1.1 \times 10^{20}$ $cm^{-3}$ [16]. The $T_{MIT}$ of approximately 250 K for the films is much lower than that for the bulk $Ti_2O_3$ (approximately 450 K) [2–9]. The lowering of $T_{MIT}$ may originate from the large $c/a$ ratio in the films [17].

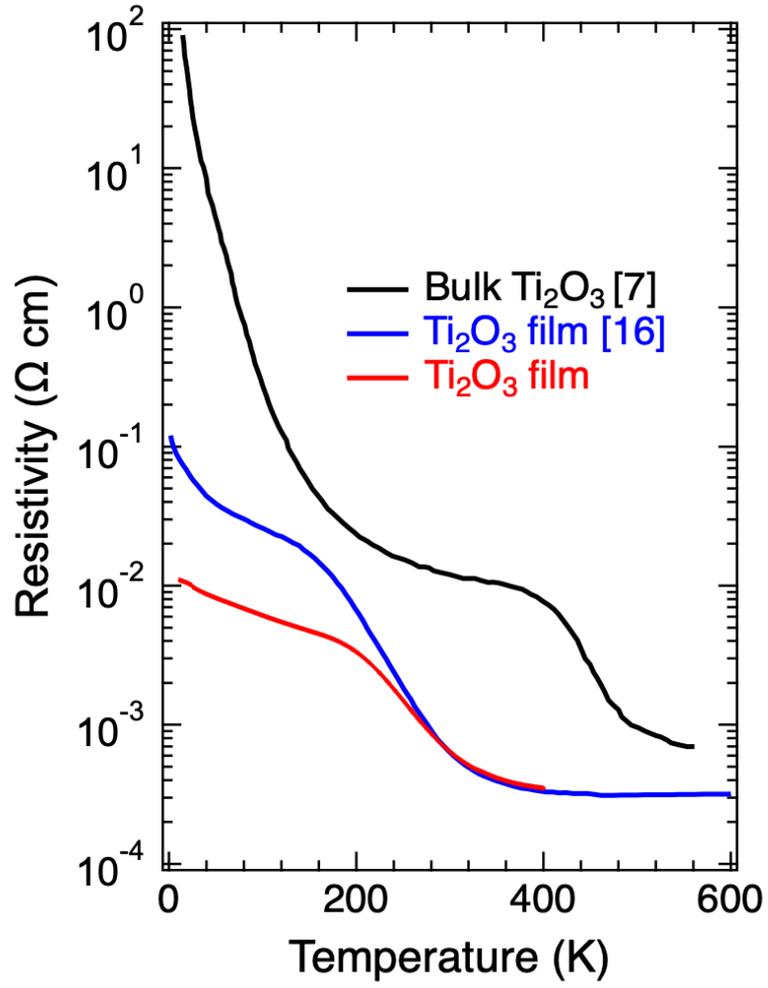

**FIG. S2.** Temperature dependence of resistivity ($\rho$–$T$) for the present $Ti_2O_3$ films. The $\rho$–$T$ curves of the $Ti_2O_3$ bulk single crystals [7] and previously reported $Ti_2O_3$ films [16] are also shown for comparison.

## 2. Band structure calculations based on DFT + $U$

### 2.1 DFT calculation details

Band-structure calculations based on DFT were conducted using the QUANTUM ESPRESSO software [53, 54]. The Perdew–Burke–Ernzerhof generalized gradient approximation (PBE-GGA) was adopted for the exchange-correlation functional [55]. The electron–ion interactions were described by ultrasoft pseudopotentials wherein atomic Ti 3$s$, 3$p$, 3$d$, and 4$s$ and O 2$s$ and 2$p$ levels were included in a valence-band state. The kinetic energy (charge density) cut-off was set to 60 (600) Ry.

The Ti and O atomic positions were optimized by the Monkhorst–Pack scheme using a $6 \times 6 \times 6$ $\boldsymbol{k}$-point grid in a self-consistent scheme. In the self-consistent calculation, the lattice constants were fixed as the experimental values ($a = 5.102$ Å, $c = 13.80$ Å; see Fig. S1). After convergence of the self-consistent calculation, the accuracy of the total energy reached less than $10^{-10}$ Ry. We then conducted a non-self-consistent calculation to draw the band dispersion using 20 $k$ points along each line. The DFT calculations were performed on the rhombohedral primitive $Ti_2O_3$ to reduce the computational cost. Thus, the band dispersion with the notation of the hexagonal Brillouin zone (BZ) was depicted by converting the rhombohedral BZ into an equivalent hexagonally shaped one [see Fig. 2(b)].

### 2.2 Possible magnetic ground states in $Ti_2O_3$ under a DFT + $U$ approximation

The possible magnetic ground states of $Ti_2O_3$ were estimated from DFT + $U$ calculations. Here, the calculations were performed assuming non-magnetic [corresponding to paramagnetic (PM)], antiferromagnetic (AFM), and ferromagnetic (FM) ground states while fixing the lattice parameters ($c/a = 2.70$) and $U$ value ($U = 2.2$ eV). The results are summarized in Table S1. The AFM states were most stable in $Ti_2O_3$, which indicates that the AFM ground state is predicted in $Ti_2O_3$ within the framework of the DFT + $U$ approximation. The local moment of Ti ions was $\pm 0.43$ $\mu_B$/Ti in the AFM states and 1.00 $\mu_B$/Ti in the most unstable FM states. The total energy difference between the PM and AFM states in $Ti_2O_3$ is considerably smaller than that between the AFM and FM states. Therefore, we have adopted the calculations for the PM ground states in $Ti_2O_3$ on the basis of the experimental fact [1].

**Table S1**. Local magnetic moment and difference in total energy relative to the ground state for paramagnetic (PM), ferromagnetic (FM), and antiferromagnetic (AFM) $Ti_2O_3$ with $c/a$ = 2.70 obtained from the DFT + $U$ calculation ($U$ = 2.2 eV).

| | Moment ($\mu_B$ / Ti) | $E - E_{AFM}$ (kJ/mol) |
|---|---|---|
| PM | – | 1.05 |
| FM | 1.00 | 11.68 |
| AFM | ± 0.43 | – |

### 2.3 Band structure of $Ti_2O_3$ obtained by DFT + $U$ calculations

Figure S3 shows the calculated band structures of $Ti_2O_3$ based on the GGA+$U$ approximation. The experimentally determined lattice constants ($c/a$ = 2.70) were used in the DFT calculations. Here, we present the results for the on-site Coulomb interaction parameter $U$ of 0 eV and 2.2 eV, and the band dispersion along representative directions ($\Gamma$–K and A–H) are used in Fig. 4, where the Fermi level ($E_F$) is shifted toward higher-binding-energy side by 150 (180 meV) to reproduce the experimental results on the hole-doped $Ti_2O_3$ films for $U$ = 0 eV (2.2 eV). As can be seen in Fig. S3 (also see the corresponding orbital projected band structures in Fig. S4), the Ti 3$d$-derived bands near $E_F$ are mainly classified into two groups based on their dispersive features, reflecting the anisotropic shape of Ti 3$d$ orbitals. One is the $a_{1g}$-derived band that is highly dispersive along the $\Gamma$–A direction (the out-of-plane direction in the present study) but less dispersive along the $\Gamma$–K direction (in-plane direction). The other is the $e_g^\pi$-derived band that shows less dispersive features along the out-of-plane direction but are highly dispersive along the in-plane directions. Meanwhile, the dispersive features along the A–H direction originate from the strong hybridization between the $a_{1g}$ states and $e_g^\pi$ and $e_g^{\,\sigma}$ states. The $e_g^\pi$-derived band forms an electron pocket(s) at the $\Gamma$ point, while the $a_{1g}$-derived band forms a hole pocket(s) at the A point for $U$ = 0 eV or at the midpoint of the $\Gamma$–A line for $U$ = 2.2 eV. The slight overlap between the two pockets makes intrinsic $Ti_2O_3$ a compensated semimetal.

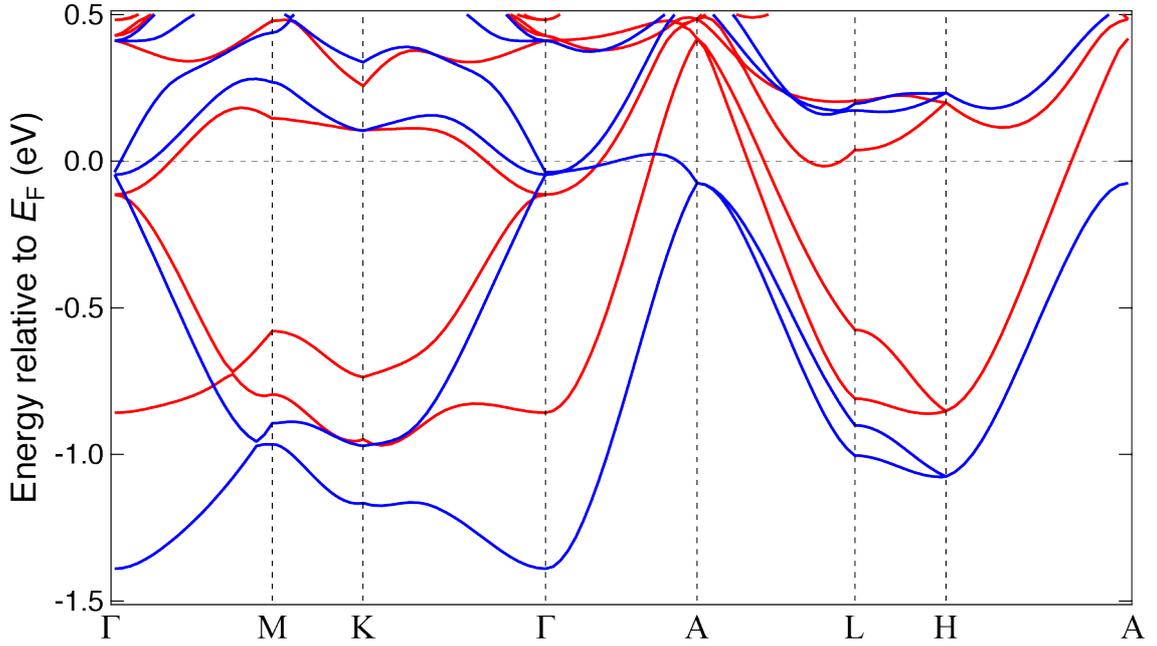

**FIG. S3.** Band structure of Ti$_2$O$_3$ obtained by the DFT + $U$ calculations. The results for $U = 0$ and 2.2 eV are given as red and blue lines, respectively.

## 2.4 Orbital projected band dispersions of Ti$_2$O$_3$

Figure S4 shows the orbital projected band dispersions of Ti$_2$O$_3$ obtained from the DFT + $U$ calculation with parameters of $c/a = 2.70$ and $U = 2.2$ eV. Note that the band dispersion itself is the same as that shown in Fig. S3. As can be seen in Fig. S4, the band structures around $E_F$ consist of the $e_g^\pi$- and $a_{1g}$-derived states. Note that the results on the orbital projection are almost similar to those previously reported [10].

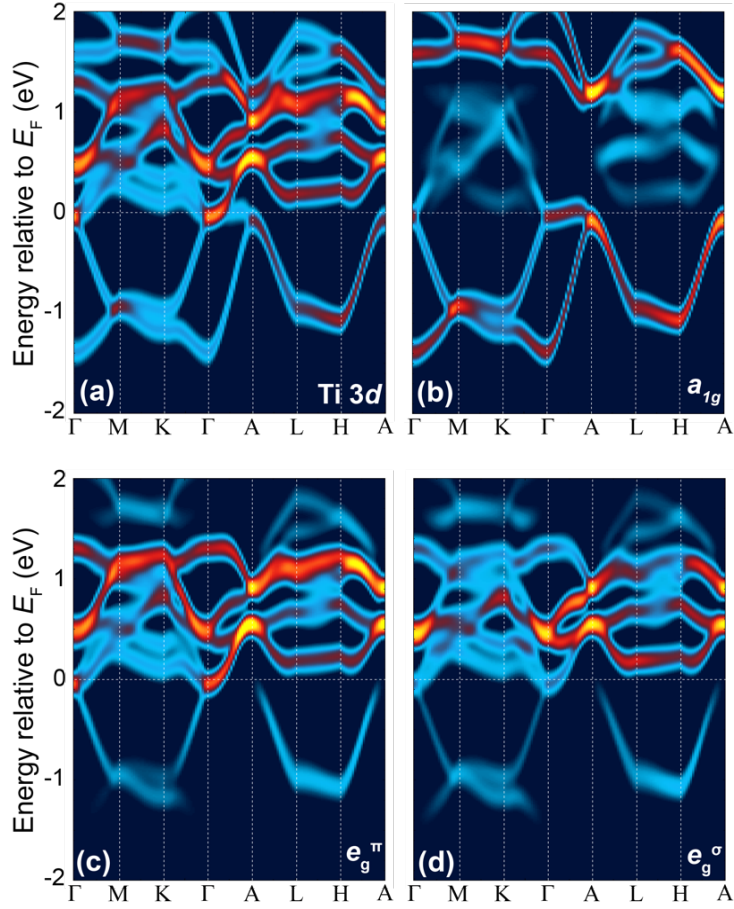

**FIG. S4.** The Ti 3$d$ orbital contribution in the band structures of Ti$_2$O$_3$ calculated by the DFT + $U$ method with $c/a$ = 2.70 and $U$ = 2.2 eV.   The contributions of the total Ti 3$d$ (a), $a_{1g}$ (b), $e_g^\pi$ (c), and $e_g^\sigma$ (d) components for the band structures are plotted by color scale.

## 2.5 $U$ dependence of the Ti$_2$O$_3$ band structure

Figure S5 shows the $U$ dependence of the band structure of Ti$_2$O$_3$ with a fixed $c/a$ ratio of 2.70. As shown in Fig. S5, Ti$_2$O$_3$ exhibits a transition from semimetal to semiconductor.   With increasing $U$, the energy separation between the $a_{1g}$ and $e_g^\pi$ states becomes large; the $a_{1g}$-derived band is pushed down, whereas the $e_g^\pi$-derived band is pushed up.   As a result, the electron and hole pockets at the Γ and A points simultaneously become smaller, and eventually a tiny energy gap opens at $U \geq 2.5$ eV.   Accordingly, the occupancy of the lowest-lying $a_{1g}$ state increases with increasing $U$.   It is also worth noting that the increment of $U$ causes narrowing of the $a_{1g}$-derived band dispersion along the A–H direction and widening of the $e_g^\pi$-derived band dispersion along the Γ–K direction.   Therefore, the bandwidth is a good indicator of the validity of the $U$ value.

The semimetallic ground states of $Ti_2O_3$ for $U = 0$–2.2 eV originate from the slight overlap in energy between the $e_g^{\pi}$- and $a_{1g}$-derived bands near $E_F$.   The increment of $U$ initially appears to reduce the overlapping.   However, a closer inspection of the Fermi surface (FS) reveals a significant change in the FS topology.   In the range of $U = 0$–2.0 eV, the electron pocket at the Γ point and hole pocket at the A point make $Ti_2O_3$ a semimetal.   However, for $U = 2.2$ eV, the electron pockets are formed at both the Γ and A points and are compensated by the hole pocket at the midpoint of the Γ–A line.   The results suggest a delicate balance between the energy of the $e_g^{\pi}$ and $a_{1g}$ states causes a change in the FS topology of $Ti_2O_3$ and consequently a significant change in the character of the conduction carriers.

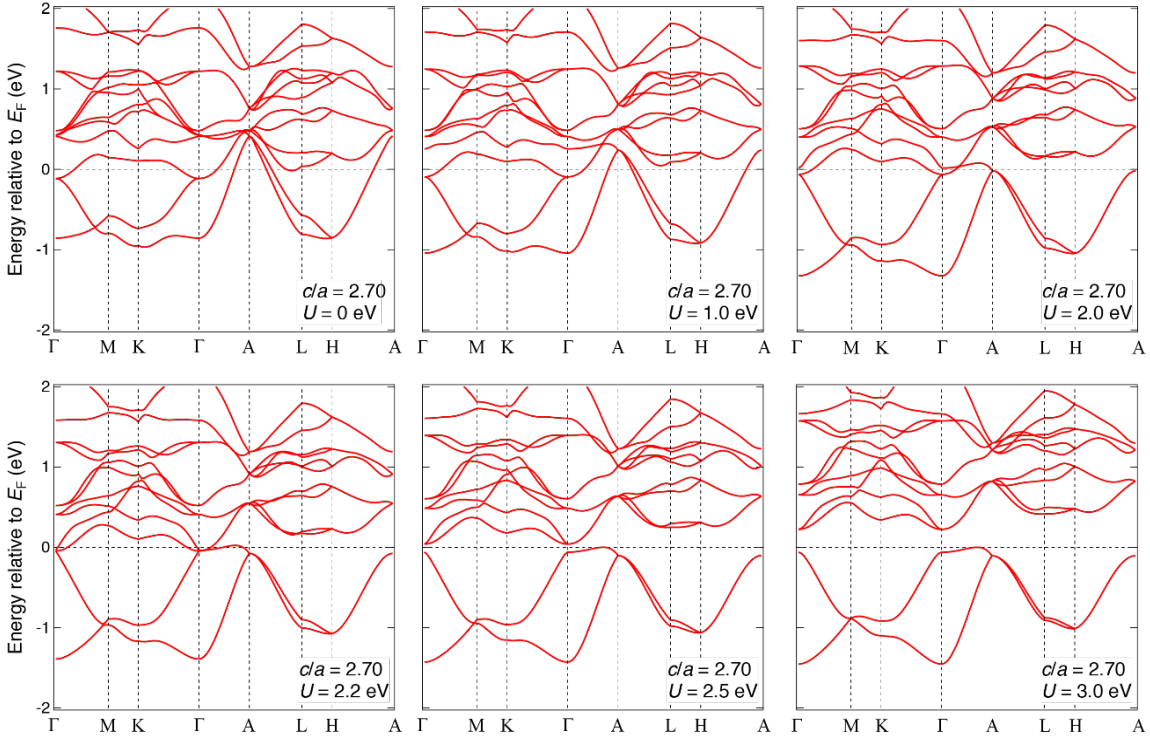

**FIG. S5.** Band structure of $Ti_2O_3$ with a $c/a$ ratio of 2.70 calculated based on the GGA + $U$ approximation.   $U$ is varied from 0 to 3.0 eV.   Note that the results for $U = 0$ and 2.2 eV are the same as those in Fig. S3.

### 2.6 *U* dependence of the band structure of Ti₂O₃ with a *c/a* ratio of 2.639

Figure S6 shows the $U$ dependence of the band structure of Ti$_2$O$_3$ with a fixed *c/a* ratio of 2.639 by varying $U$ from 0 to 3.0 eV.   The overall band structures and their $U$ dependence are quite similar to those of Ti$_2$O$_3$ with a *c/a* ratio of 2.70 (see Fig. S5).   A crucial difference is the critical $U$ value for the MIT.   In Ti$_2$O$_3$ with the *c/a* ratio of 2.639, the MIT occurs at the critical $U$ value of 2.0 eV, whereas Ti$_2$O$_3$ with the *c/a* ratio of 2.70 exhibited the MIT at the critical $U$ value of 2.5 eV (see Fig. S5).   This difference in the critical $U$ values suggests that the electronic structures of Ti$_2$O$_3$ also strongly depend on lattice deformation, as shown in Fig. S7.

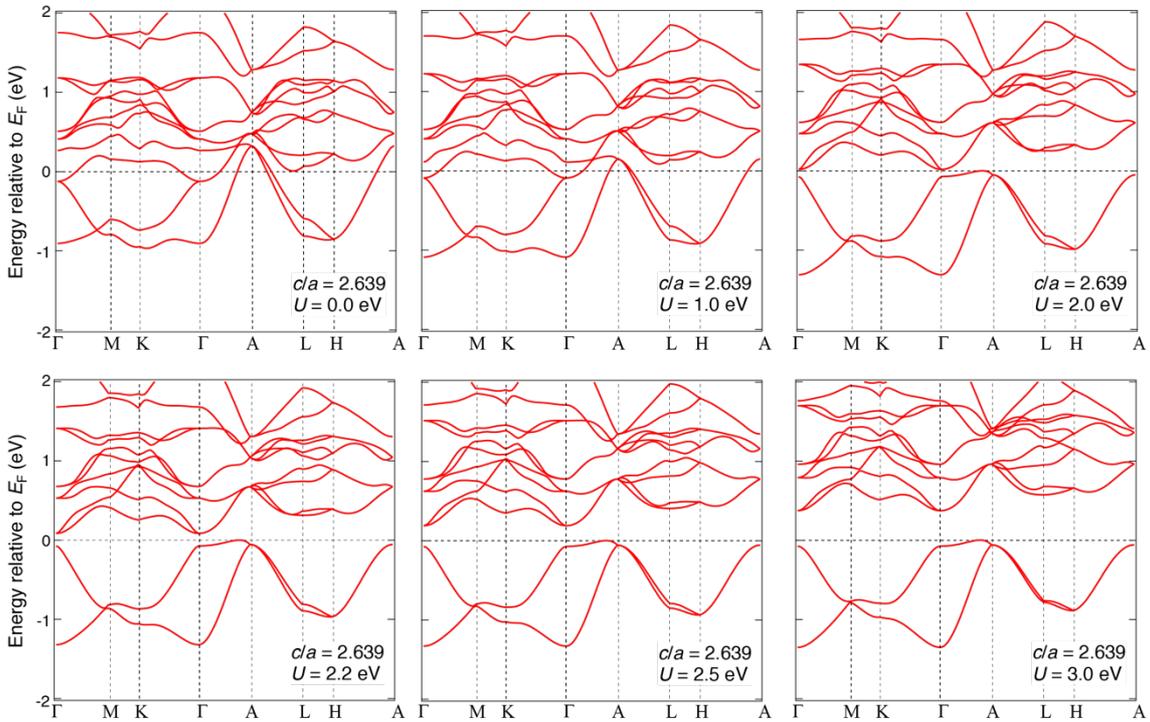

**FIG. S6.** Band structures of Ti$_2$O$_3$ with a *c/a* ratio of 2.639 calculated based on the GGA + $U$ approximation.   $U$ is varied from 0 to 3.0 eV.

### 2.7 *c/a* ratio dependence of the Ti₂O₃ band structure

To investigate the effect of lattice deformation on the electronic band structure of $Ti_2O_3$, band-structure calculations were conducted by varying the $c/a$ ratio while keeping the $U$ value fixed to 2.2 eV.    Figure S7 shows the obtained $c/a$-ratio dependence of the band structure of $Ti_2O_3$. With increasing the $c/a$ ratio, the energy separation between the $a_{1g}$ and $e_g^\pi$ states becomes smaller; the $a_{1g}$-derived band is pushed up, whereas the $e_g^\pi$-derived band is pushed down.    As a result, electron pockets appear at the $\Gamma$ point, and simultaneously, a hole pocket appears at the midpoint of the $\Gamma$–A line for a critical $c/a$ ratio of 2.68.    The semimetallic ground states of $Ti_2O_3$ for $c/a \geq 2.68$ originate from the slight overlap in energy between the $e_g^\pi$- and $a_{1g}$-derived bands near $E_F$.    Interestingly, the $c/a$-ratio dependence of these band structures (with increasing $c/a$) appears to be similar to the $U$ dependence (with decreasing $U$) as shown in Figs. S5 and S6. These results demonstrate that the band structures of $Ti_2O_3$ strongly depend on both the parameters of the $c/a$ ratio and $U$.    Furthermore, the resultant delicate balance between the energy of the $e_g^\pi$ and $a_{1g}$ states causes a change in the FS topology of $Ti_2O_3$, and consequently, a significant change in the character of the conduction carriers.

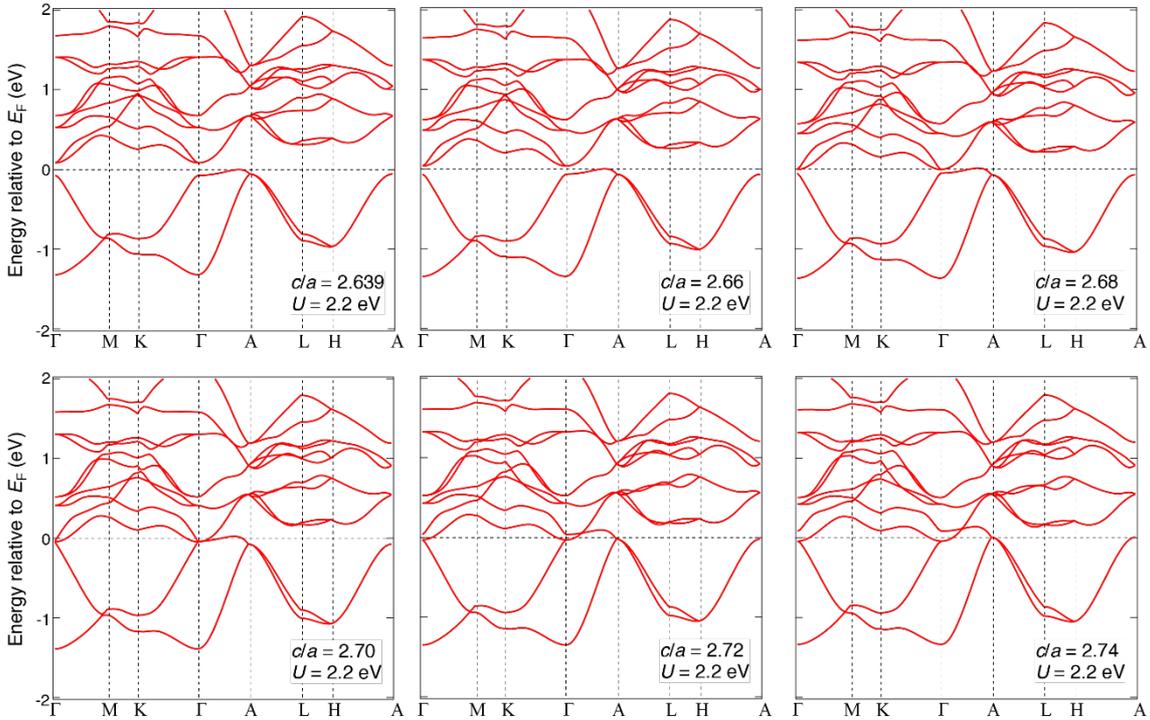

**FIG. S7.** Band structure of $Ti_2O_3$ calculated with a fixed $U$ value of 2.2 eV based on the GGA + $U$ approximation.    The $c/a$ ratio is varied from 2.639 to 2.74.

### 2.8 Fermi surface of Ti$_2$O$_3$

Figure S8 shows the Fermi surface (FS) of Ti$_2$O$_3$ corresponding to the band-structure calculation shown in Figs. 3 and 4 ($c/a$ = 2.70 and $U$ = 2.2 eV).    Here, the Fermi level is shifted toward the higher-binding-energy-side for comparison with the ARPES results on the hole-doped Ti$_2$O$_3$ films, which is discussed in the main text.    The cross-sectional FSs in the Γ–A–H–K plane [Fig. S8(b)], Γ–K–M plane [Fig. S8(c)], and A–H–L plane [Fig. S8(d)] are shown for comparison with the ARPES results.    As shown in Fig. S9, the FSs determined by ARPES are well reproduced by the DFT + $U$ calculation, demonstrating that the determination of $U$ = 2.2 eV is reasonable to reproduce the electronic structures of Ti$_2$O$_3$ in the framework of the present DFT + $U$ approximation.

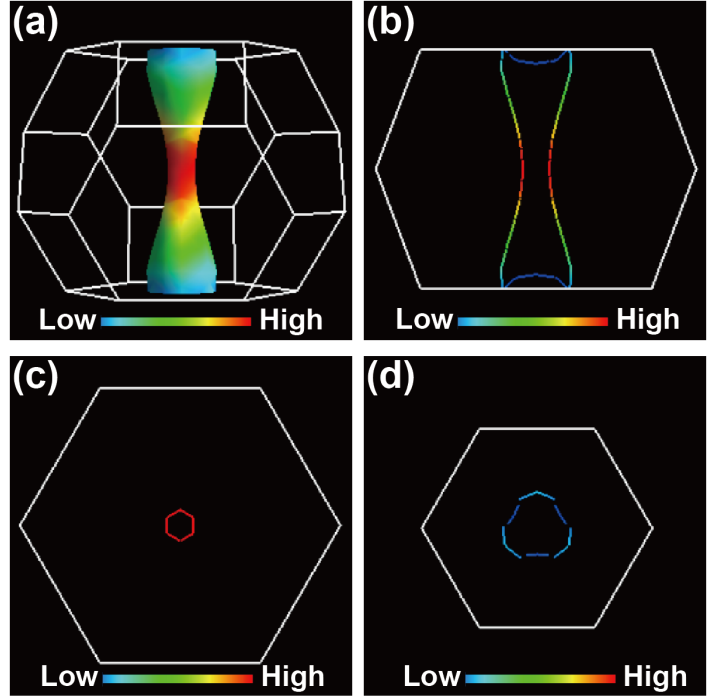

**FIG. S8.** (a) Fermi surfaces of Ti$_2$O$_3$ obtained from the DFT + $U$ calculation ($c/a$ = 2.70, $U$ = 2.2 eV).    The white lines indicate the first BZ of rhombohedral Ti$_2$O$_3$.    For comparison with the ARPES results, the Fermi level is shifted toward the higher-binding-energy side as in the same manner described in the text.    (b–d) Fermi surfaces cut at representative planes. (b), (c), and (d) correspond to the Γ–A–H–K, Γ–K–M, and A–H–L planes in the hexagonally-shaped BZ of Ti$_2$O$_3$, respectively.    The weight of the Ti 3$d$ components for the FS is shown by color scale.    The FSs are visualized using the FermiSurfer software [S1].

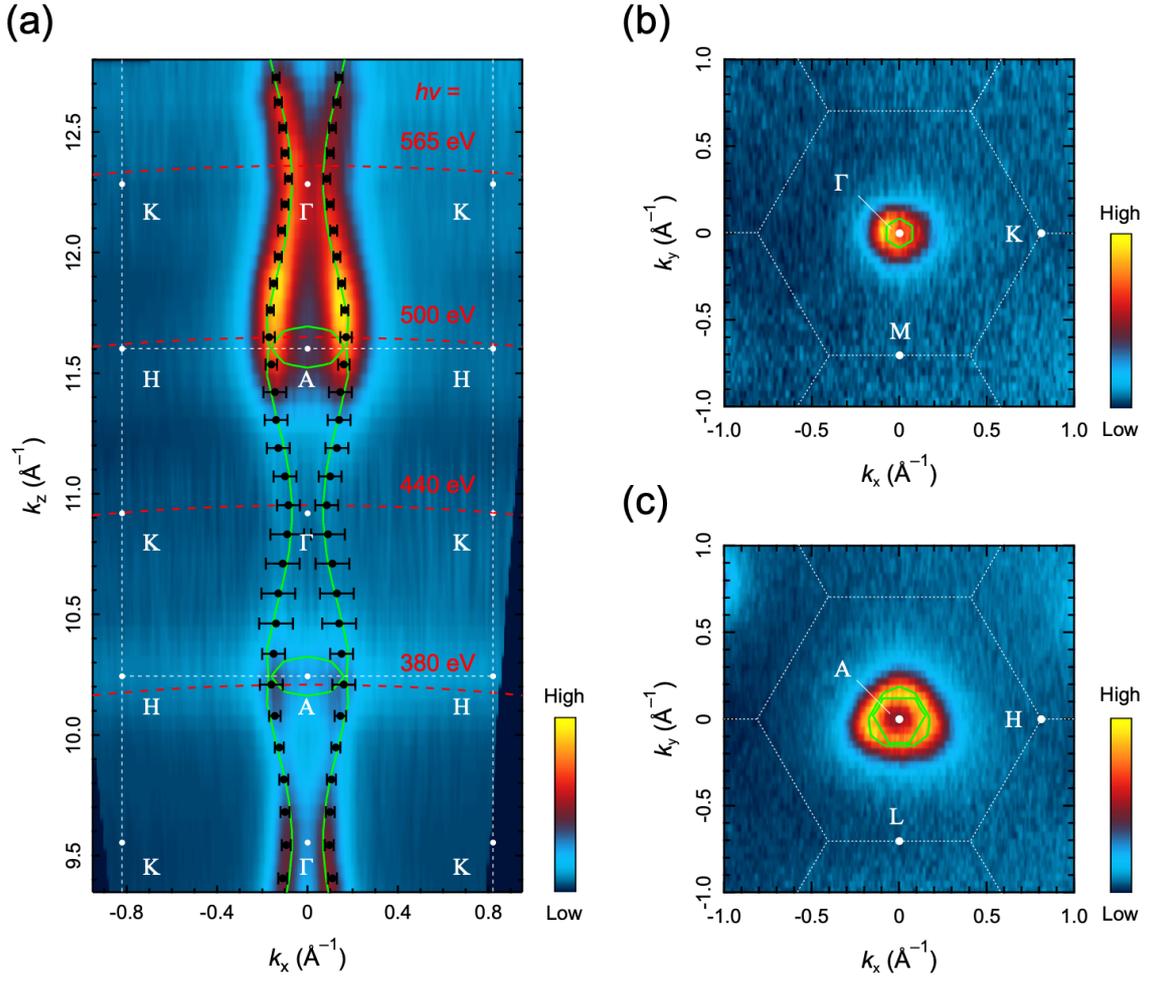

**FIG. S9.** Comparison of the Fermi surfaces between the ARPES results and DFT + $U$ calculations on the Γ–A–H–K (a), Γ–K–M (b), and A–H–L (c) planes in the hexagonally-shaped BZ of $Ti_2O_3$. The excellent agreement indicates the essential validity of our band-structure calculation (indicated by green lines) and ARPES analysis procedures.

## 2.9 Benchmark for evaluating the appropriate *U* value

As described in the text, we used $U = 2.2$ eV in the DFT + $U$ calculation for comparison with the experimental results (Figs. 3 and 4). This is because the DFT calculation with $U = 2.2$ eV well reproduced the insulating phase of *bulk* $Ti_2O_3$, which exhibits a small energy gap of approximately 100 meV at a *c/a* ratio of 2.639 [17]. The present DFT + $U$ calculation with $U = 2.2$ eV well reproduces the energy gap in *bulk* $Ti_2O_3$ (see Figs. S6 and S7). The application of the same *U* value to the films is reasonable because the on-site Coulomb repulsion is not sensitive to changes in the environment of Ti ions owing to the narrow spatial distribution of Ti 3*d* electrons. Therefore, in our calculation for the $Ti_2O_3$ *films* shown in Figs. 3 and 4 (also Figs. S3–S5, S8, and S9), we used $U = 2.2$ eV and the *c/a* ratio of 2.70, which was determined from X-ray diffraction measurements at room temperature (see the reciprocal space map in Fig. S1). The suitability of the parameters is confirmed by the fact that the ARPES results are well reproduced by the calculation, especially the Fermi momentum $k_F$ (Fig. 4) and FS topology (Fig. S9).

To further confirm the validity of $U = 2.2$ eV, we evaluated the suitability of the *U* value as follow. When *U* is not uniquely determined by comparing the band dispersion, as in the present case, the criterion for determining the appropriate *U* is the agreement in the $k_F$ values between the DFT + $U$ calculations and ARPES results. Thus, we utilized the $k_F$ value in the same manner as described in the manuscript. We shifted the Fermi level in the DFT + $U$ calculations with different *U* values (Fig. S5) by $\Delta E$ toward the higher-binding-energy side to reproduce the $k_F$ at the Γ point, reflecting the hole-doped nature of the $Ti_2O_3$ films. Next, we judged the validity of *U* by evaluating the agreement of the $k_F$ at the A point.

Figure S10 shows a series of the comparison between the experimental band structure and DFT + $U$ calculations with different *U* values. In the plot for the difference in the $k_F$ at the A point between the ARPES results and DFT + $U$ calculations (Fig. S11), it is clear that the best match is obtained when $U = 2.2$ eV, although appropriate *U* values seem to be in the range of $U = 2.2$–2.4 eV by considering the experimental error. Based on the evaluation, we conclude that $U = 2.2$ eV is the most appropriate value for describing the band structure of $Ti_2O_3$ thin films.

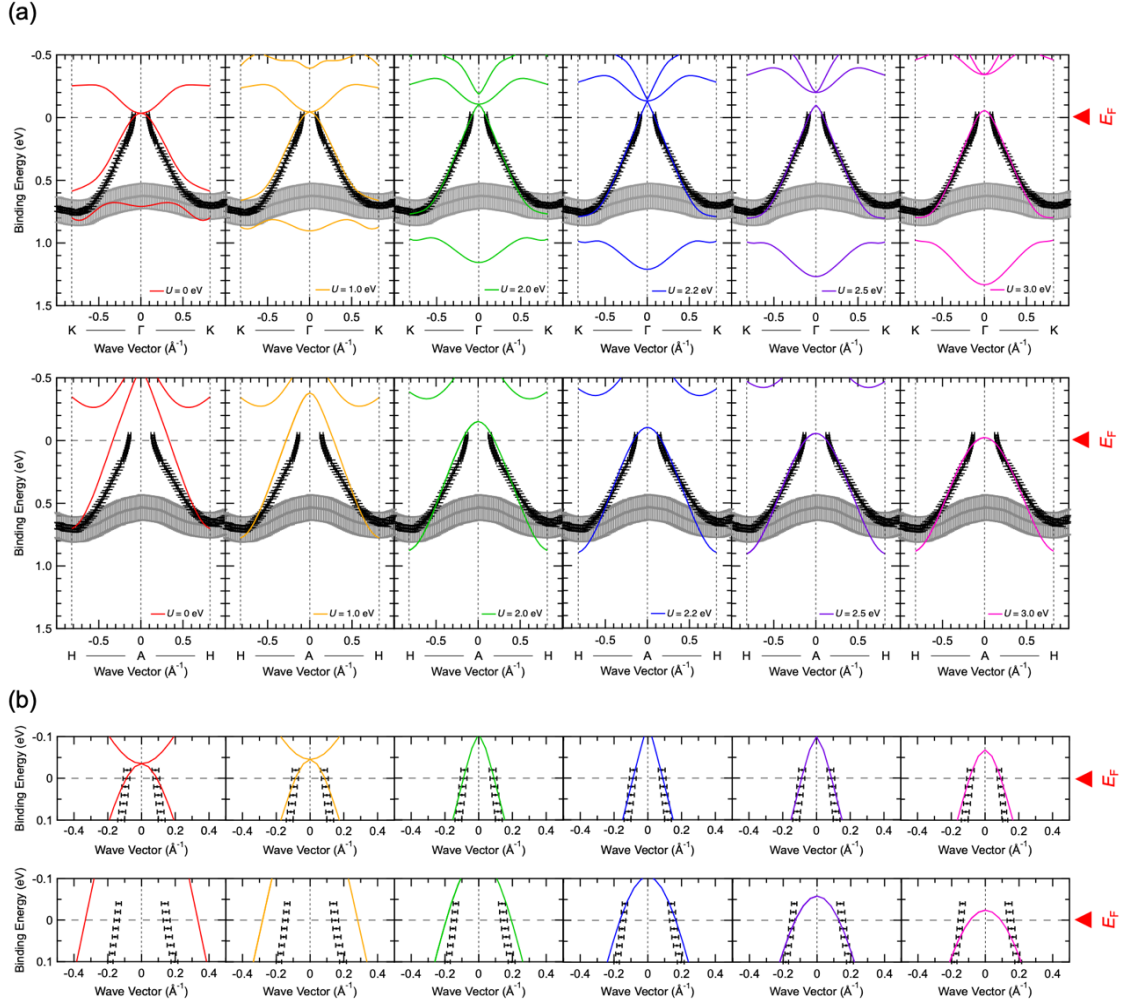

**FIG. S10.** Comparison of the experimental band structure and DFT + $U$ calculation (a) in the energy range from 1.5 to –0.5 eV and (b) in the magnified scale near $E_\mathrm{F}$ along the Γ–K (upper panels) and A–H (lower panels) directions. The $U$ value is varied from 0 to 3.0 eV with a constant $c/a$ ratio of 2.70. The data markers indicate the Ti 3$d$ bands determined by ARPES (same as those in Fig. 4), while the colored lines correspond to the results of the DFT + $U$ calculations. The DFT + $U$ results with $U = 0$ and 2.2 eV are the same as those used in Fig. 4. Note that the Fermi level in the DFT + $U$ calculations is shifted toward the higher-binding-energy side to reproduce the $k_\mathrm{F}$ at the Γ point. It is evident that the best match is obtained at $U = 2.2$ eV.

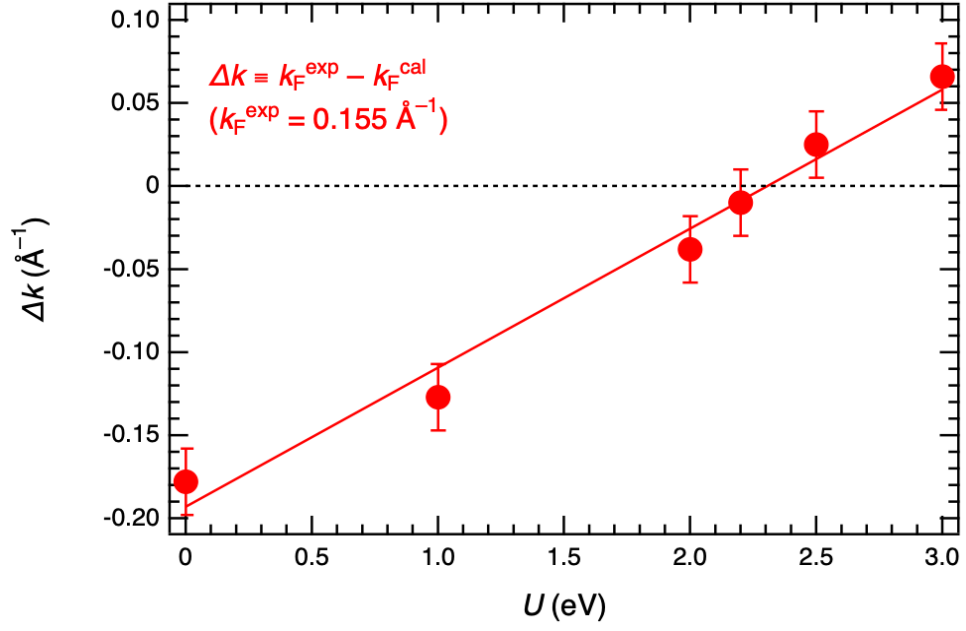

**FIG. S11.** Plot of differences in the $k_F$ ($\Delta k$) between the ARPES experiments ($k_F^{exp}$) and DFT + $U$ calculations ($k_F^{cal}$) at the A point as a function of $U$.

## 3. Analysis details of ARPES data

### 3.1 Out-of-plane FS mapping

Figure S12(a) shows the out-of-plane FS map of the $Ti_2O_3$ films obtained from normal-emission ARPES measurements in the Γ–A–H–K emission plane. Normal-emission (photon-energy dependent) ARPES images were taken along the $k_x$ slice with changing the photon energy ($hv$) from 300 to 600 eV in 10 eV increments. Reflecting the hole-doped nature of the films, a meandering FS that follows the periodicity of the bulk BZ is clearly observed along the BZ center line. The cross-section of the FS is largest at the A points (corresponding to $hv$ = 500 and 380 eV), monotonically decreases away from the A points, and is smallest at the Γ points (corresponding to $hv$ = 560 and 440 eV). From this periodicity, the inner potential was determined to be approximately 21.8 eV.

### 3.2 Band dispersion along the Γ–A direction

Figure S12(b) shows the ARPES image along the Γ–A direction, which is drawn by picking up the normal-emission ARPES spectra from the photon-energy dependent ARPES images. A series of the normal-emission ARPES spectra is shown in Fig. S12(c). As can be seen in Fig. S12(b), some flat bands are only dimly observed in the ARPES image; broad and nondispersive quasi-localized states are located at approximately 0.6 eV, while flat states are located just below $E_F$ for the bottom and top Γ points. The former may originate from the surface-derived states characteristic of corundum-type conductive oxides [42], while the latter may originate from the angle-integrated effects of the hole band that forms a closed hole surface in the proximity of the zone center line [see Fig. S12(a)]. However, a closer inspection in the energy range of 460–500 eV reveals the existence of a highly dispersive feature near $E_F$, which approaches $E_F$ with increasing photon energy and may cross $E_F$ around the A point, suggesting the formation of an additional small hole FS around the A point. In comparison with the DFT calculation [10] (also see Fig. S4), the dispersive band is attributable to the $a_{1g}$-derived band, which disperses highly along the Γ–A direction (out-of-plane direction), reflecting the anisotropic shape of the $a_{1g}$ orbital. Meanwhile, the bottom of the band around the Γ point is unclear owing to the overlap with the surface-derived nondispersive quasi-localized states and certain matrix-element effects.

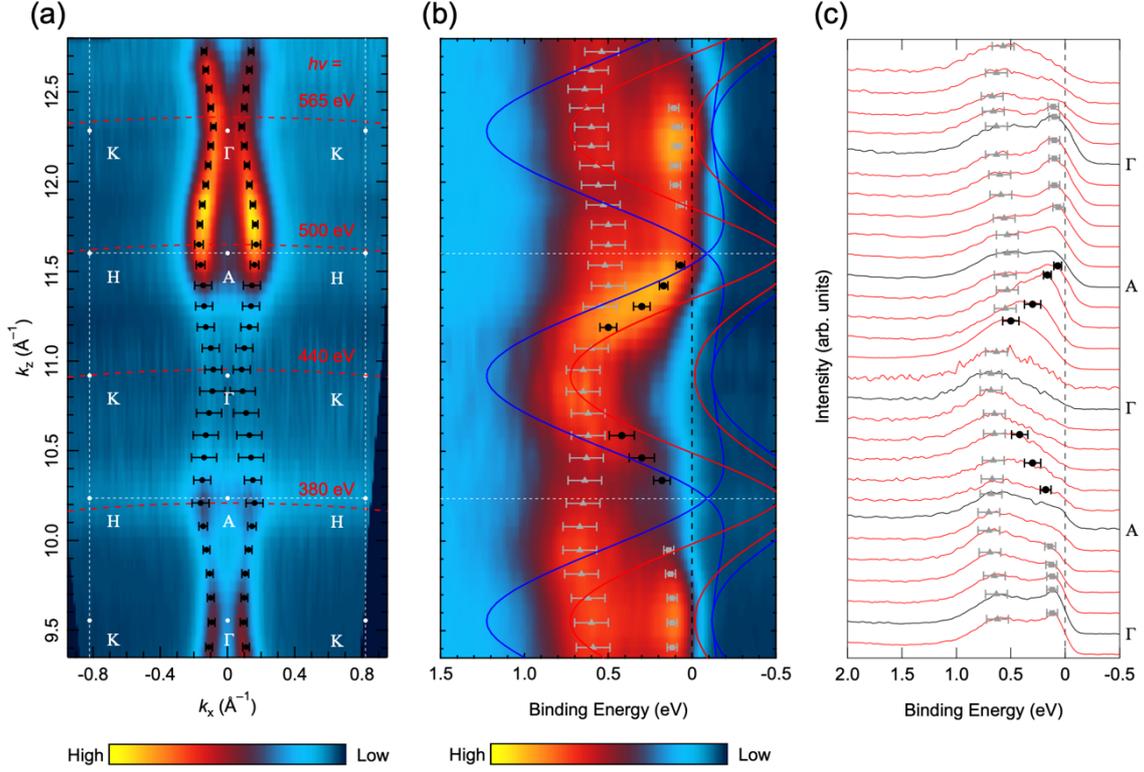

**FIG. S12.** (a) Out-of-plane FS map for the $Ti_2O_3$ films determined by photon-energy dependent ARPES images in the Γ–A–H–K emission plane. The FS map was obtained by plotting the ARPES intensity within the energy window of $E_F\pm50$ meV. The data markers indicate the Fermi momentum ($k_F$) determined by fitting the momentum distribution curves (MDCs) of the corresponding ARPES image. The white dotted lines indicate the hexagonal BZ, and the red dotted arc lines represent the $\mathbf{k}$ paths passing through the A points (at photon energies of 500 eV and 380 eV) and the Γ points (at photon energies of 565 eV and 400 eV). (b) Experimental band structure along the Γ–A direction obtained by normal-emission ARPES. Black markers indicate the positions of the $a_{1g}$-derived band. Gray triangle and square markers indicate the positions of the surface-derived quasi-localized states and artifacts originating from the angle-integrated effects, respectively. For comparison purposes, the calculated band dispersions with $U = 0$ eV and 2.2 eV are overlayed as red and blue lines, respectively. (c) Corresponding normal-emission ARPES spectra along the Γ–A direction. The spectra were picked up from the photon-energy dependent ARPES images shown in Fig. S13.

### 3.3 ARPES images taken at different photon energies

The existence of the highly dispersive $a_{1g}$-derived band along the Γ–A direction is further confirmed by the photon-energy dependent ARPES images. Figure S13 shows representative ARPES images taken at different photon energies from 440 to 510 eV. In addition to the highly dispersive $e_g^{\pi}$-derived band, which forms the open hole FS along the BZ center line [Fig. S12(a)], another feature is observed in the ARPES images for $h\nu$ = 460–490 eV. The additional band is attributed to the $a_{1g}$-derived band along the Γ–A direction in comparison with the DFT calculation (Fig. S4) [10]. The $a_{1g}$-derived band approaches $E_F$ with increasing $h\nu$ and may cross $E_F$ around the A point, forming a closed FS at the A point as predicted from the DFT calculation (see Fig. 4 and Figs. S8 and S9).

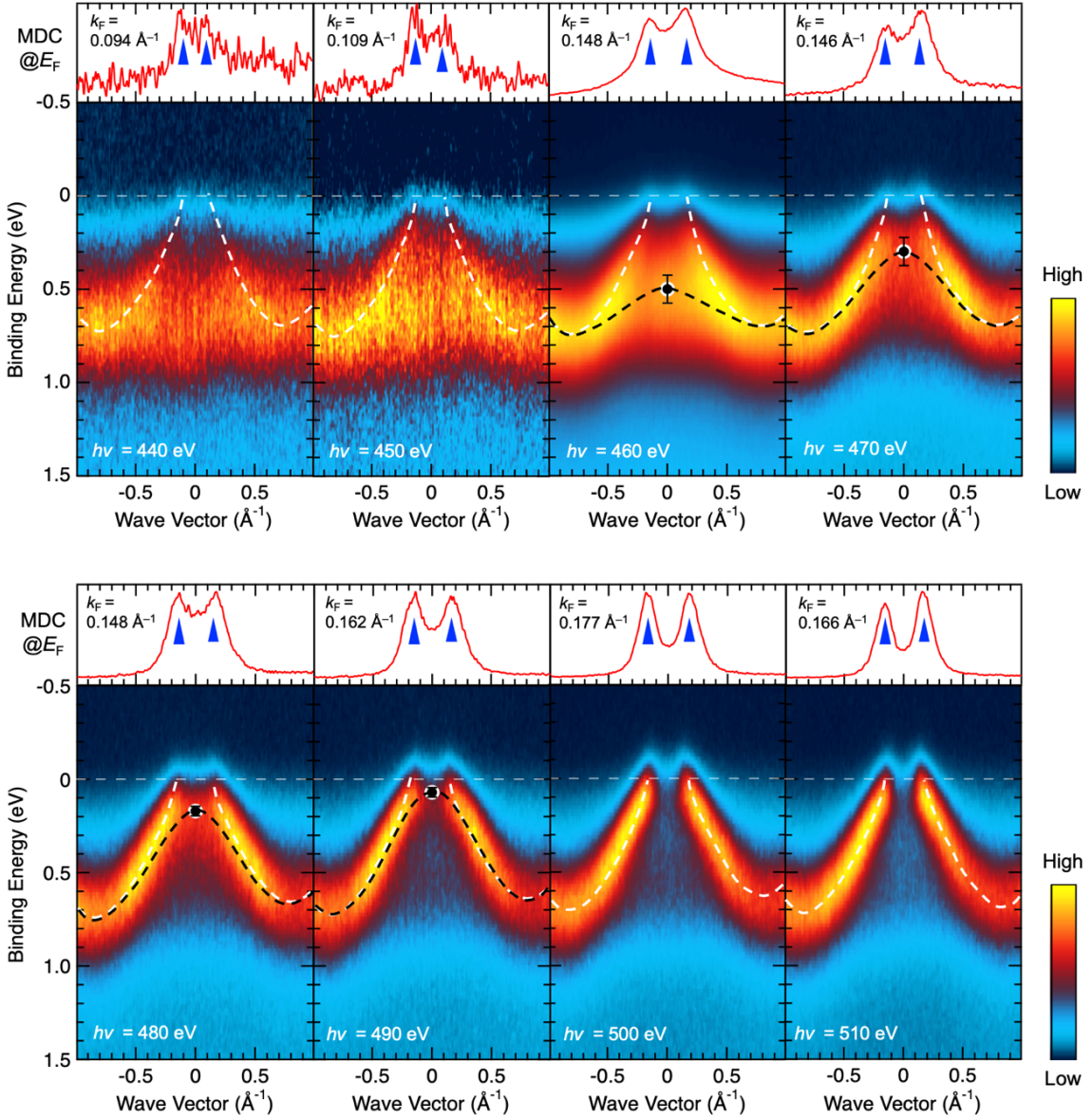

**FIG. S13.** Photon-energy dependence of ARPES images taken in the photon energy range of 440–510 eV. The white and black dotted lines are visual guides for the band dispersions of $e_g^\pi$-derived and $a_{1g}$-derived bands, respectively. The MDCs at $E_F$ are also shown at the top panel of each ARPES image. The blue triangles indicate the position of $k_F$. The outer $e_g^\pi$-derived band forms a hole pocket at the zone center irrespective of $h\nu$ ($k_z$), whereas the top of the $a_{1g}$-derived band located below $E_F$ at $h\nu = 460$ eV approaches $E_F$ with increasing $h\nu$ and eventually crosses $E_F$ (degenerate with the $e_g^\pi$-derived band) at $h\nu = 500$ eV (the A point). The data markers correspond to those in Figs. S12(b) and (c).

### 3.4 Fermi surface mapping for Ti₂O₃ films in the Γ–K–M and A–H–L planes

Figure S14 shows the FS maps taken at different photon energies for the A–H–L (taken at $h\nu =$ 500 and 382 eV) and Γ–K–M (taken at $h\nu =$ 565 and 440 eV) emission planes. Note that Figs. S14(a) and (b) are the same as the left and right sides of Fig. 2(d), respectively. The observed triangular-like FS around the A point in Fig. S14(a) is reversed at the other A point of one point below [Fig. S14(c)], reflecting the original trigonal symmetry of Ti₂O₃. Together with the overall threefold intensity pattern of the observed FS, the results provide evidence that the present ARPES measurements observe the "bulk" electronic structures of Ti₂O₃.

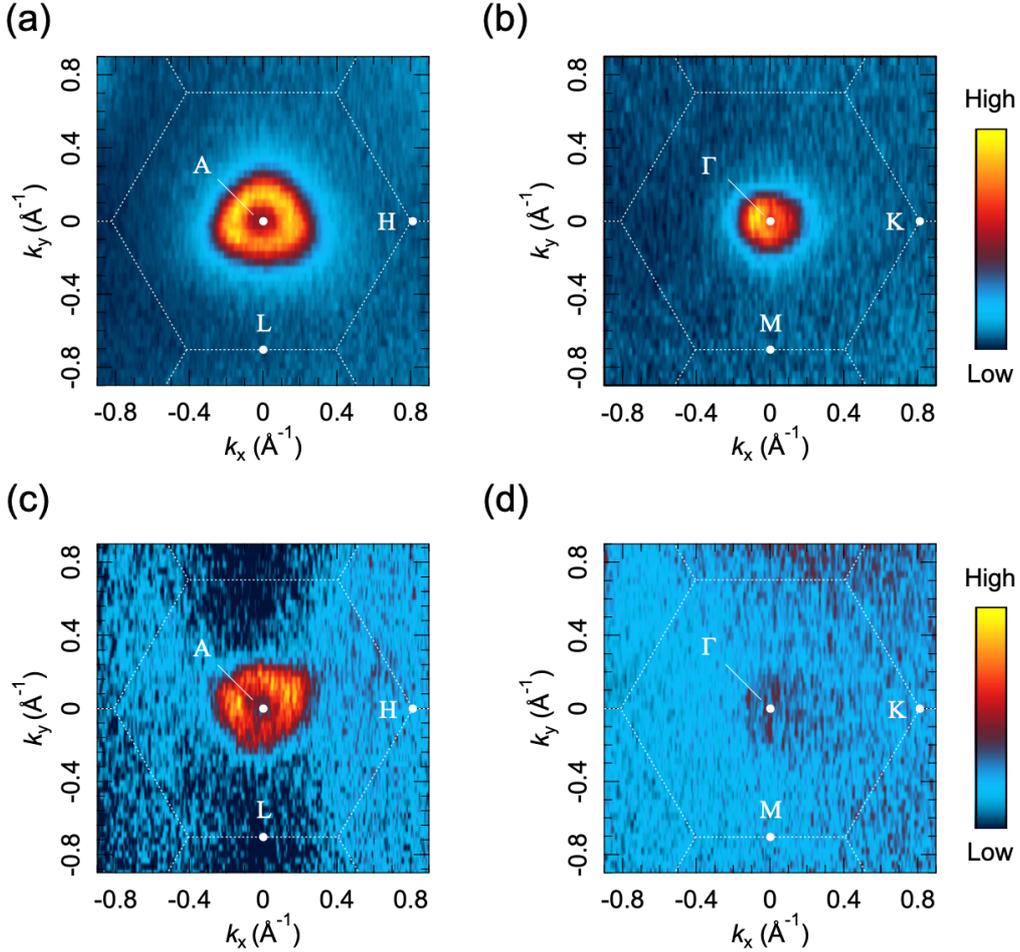

**FIG. S14.** FS maps of the Ti₂O₃ films acquired at constant photon energies of (a) 500 eV, (b) 565 eV, (c) 382 eV, and (d) 440 eV by changing emission angles. The hexagonal BZs are overlayed as the white dotted lines on the FS maps. The FS maps were obtained by plotting the ARPES intensity within the energy window of $E_F \pm 50$ meV. The much weaker intensity at the A and Γ points of one point below (382 eV and 440 eV) is due to the matrix-element effects (see Fig. S12).

### 3.5 Energy distribution curves along the A–H and Γ–K directions

Figures S15(a) and (b) show the energy distribution curves (EDCs) of the Ti$_2$O$_3$ films obtained at photon energies of 500 eV (corresponding to the A–H line) and 565 eV (corresponding to the Γ–K line), respectively. Dispersive features are clearly observable in the EDCs. Note that the seeming "peak" structure at the Γ point is due to the angle-integrated effect and close proximity of the hole FS with high intensity.

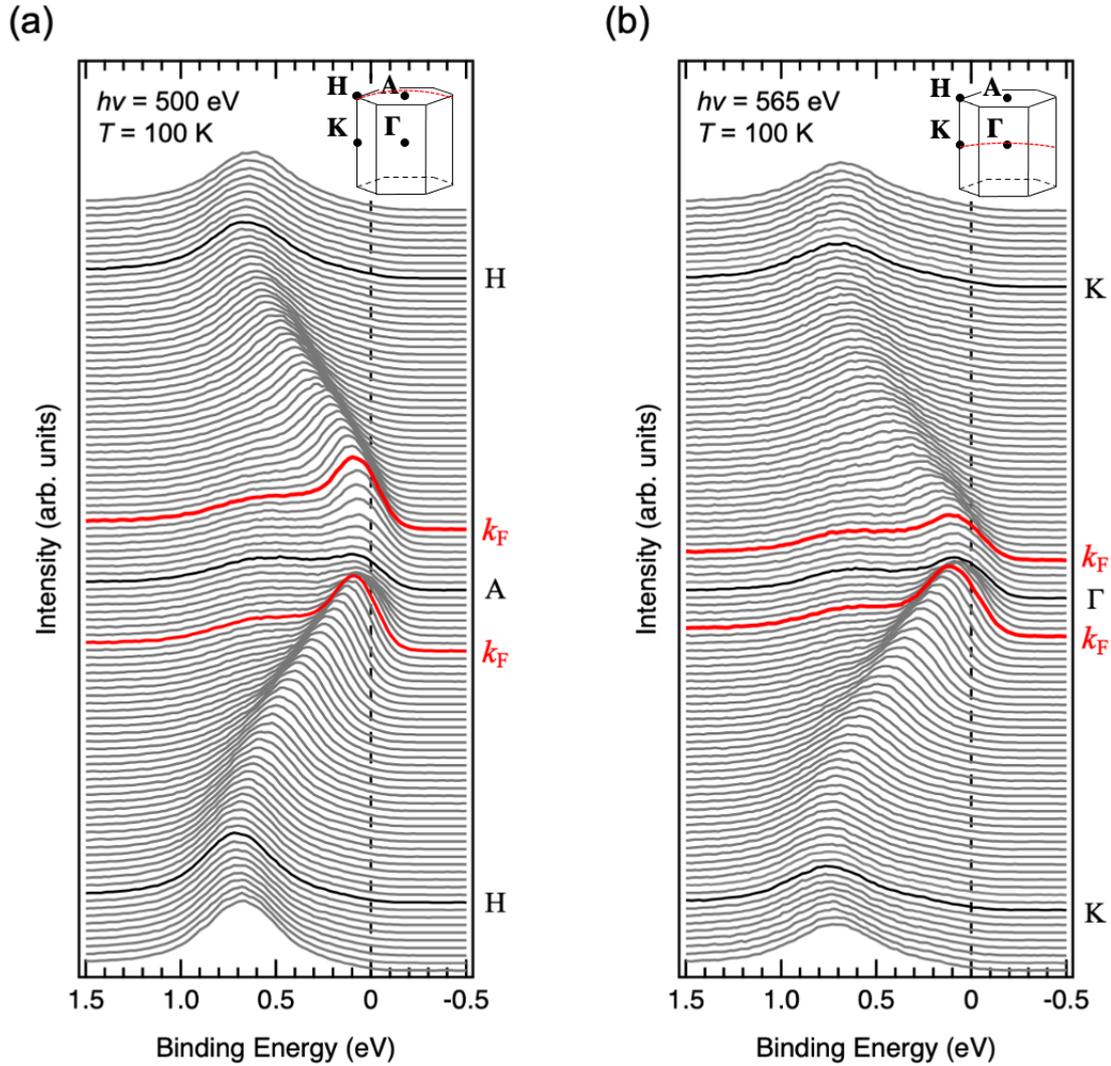

**FIG. S15.** EDCs (ARPES spectra) of the Ti$_2$O$_3$ films taken along the (a) A–H and (b) Γ–K directions. The corresponding measurement lines in the BZ are indicated by the red dotted arc lines in each inset.

## Reference

[S1] M. Kawamura, "FermiSurfer: Fermi-surface viewer providing multiple representation schemes" Comp. Phys. Commun. **239**, 197–203 (2019). DOI: 10.1016/j.cpc.2019.01.017.